\documentclass[11pt,a4paper]{article}
\pdfoutput=1
\usepackage{jheppub}
\usepackage{amsmath}
\usepackage{bm}
\usepackage{times}
\usepackage{braket}
\usepackage{color,graphicx}
\usepackage[T1]{fontenc}
\usepackage[utf8]{inputenc}
\usepackage{hyperref}
\usepackage{lscape}
\usepackage[normalem]{ulem}

\definecolor{violet}{cmyk}{0,1,0,0.2}

\usepackage{xspace}
\newcommand{\MadGraph}{{\rmfamily\scshape MadGraph5\_aMC@NLO}\xspace}
\newcommand{\MadLoop}{{\rmfamily\scshape MadLoop}\xspace}
\newcommand{\MadFKS}{{\rmfamily\scshape MadFKS}\xspace}
\newcommand{\Ninja}{{\rmfamily\scshape Ninja}\xspace}

\newcommand{\FeynRules}{{\rmfamily\scshape FeynRules}\xspace}
\newcommand{\FeynArts}{{\rmfamily\scshape FeynArts}\xspace}
\newcommand{\NLOCT}{{\rmfamily\scshape NLOCT}\xspace}
\newcommand{\UFO}{{\rmfamily\scshape UFO}\xspace}

\newcommand{ \PDFLHC}{{\rmfamily\scshape PDF4LHC15}\xspace}

\title{Leptoquark toolbox for precision collider studies}

\author[a]{Ilja Dor\v sner} 
\author[b, c]{and Admir Greljo} 

\affiliation[a]{University of Split, Faculty of Electrical Engineering, Mechanical Engineering and Naval Architecture in Split (FESB), Ru\dj era Bo\v skovi\' ca 32, 21000 Split, Croatia}
\affiliation[b]{PRISMA Cluster of Excellence and Mainz Institute for Theoretical Physics, Johannes Gutenberg-Universit\"at Mainz, 55099 Mainz, Germany}
\affiliation[c]{Faculty of Science, University of Sarajevo, Zmaja od Bosne 33-35, 71000 Sarajevo, Bosnia and Herzegovina}

\emailAdd{dorsner@fesb.hr}
\emailAdd{admgrelj@uni-mainz.de} 
  
\abstract{We implement scalar and vector leptoquark (LQ) models in the universal \FeynRules output (\UFO) format assuming the Standard Model fermion content and conservation of baryon and lepton numbers. Scalar LQ implementations include next-to-leading order (NLO) QCD corrections. We report the NLO QCD inclusive cross sections in proton-proton collisions at 13\,TeV, 14\,TeV, and 27\,TeV for all on-shell LQ production processes. These comprise (i) LQ pair production ($p p \to \Phi \Phi$) and (ii) single LQ + lepton production ($p p \to \Phi \ell$) for all initial quark flavours ($u$, $d$, $s$, $c$, and $b$). Vector LQ implementation includes adjustable non-minimal QCD coupling.
We discuss several aspects of LQ searches at a hadron collider, emphasising the implications of $SU(2)$ gauge invariance, electroweak and flavour constraints, on the possible signatures. Finally, we outline the high-$p_T$ search strategy for LQs recently proposed in the literature to resolve experimental anomalies in $B$-meson decays. In this context, we stress the importance of complementarity of the three LQ related processes, namely, $p p \to \Phi \Phi$, $p p \to \Phi \ell$, and $p p \to \ell \ell$.}

\arxivnumber{1801.07641}

\begin{document}

\maketitle

\section{Introduction}
\label{sec:introduction}

Leptoquarks (LQs) are either vector or scalar fields that couple a quark to a lepton at the tree level. This feature makes them rather unique within a plethora of hypothetical particles that are being experimentally searched for. LQs are coloured objects that always reside in the (anti)fundamental representations of the $SU(3)$ part of the $SU(3) \times SU(2) \times U(1)$ gauge group of the Standard Model (SM). They can thus be copiously produced at hadron machines such as the LHC if kinematically accessible. Moreover, the fact that they decay into a quark and a lepton practically guarantees measurable signatures at modern particle detectors. 

\begin{figure}[tbp]
\centering
\includegraphics[scale=0.19]{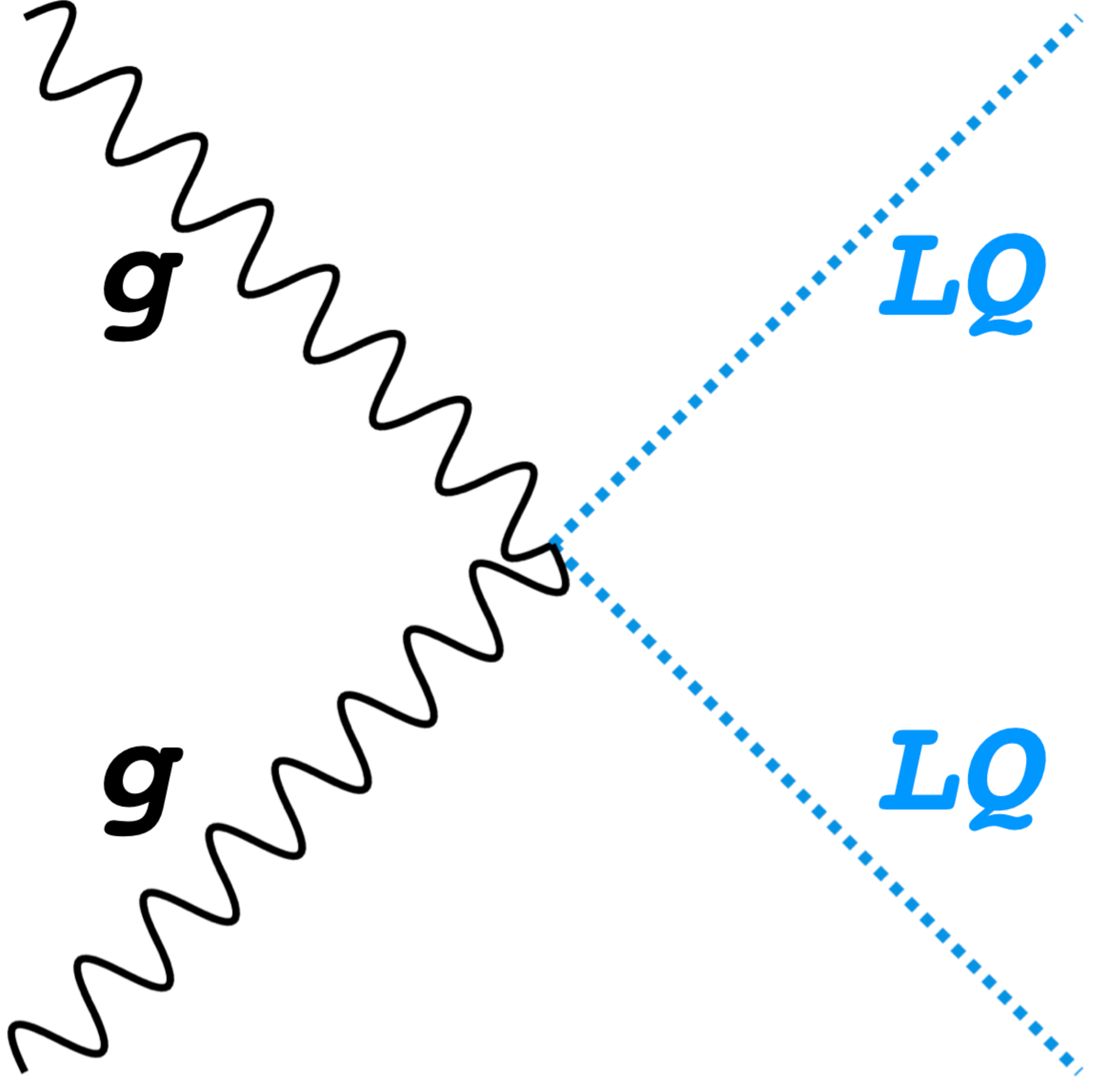} \; \; \; \includegraphics[scale=0.19]{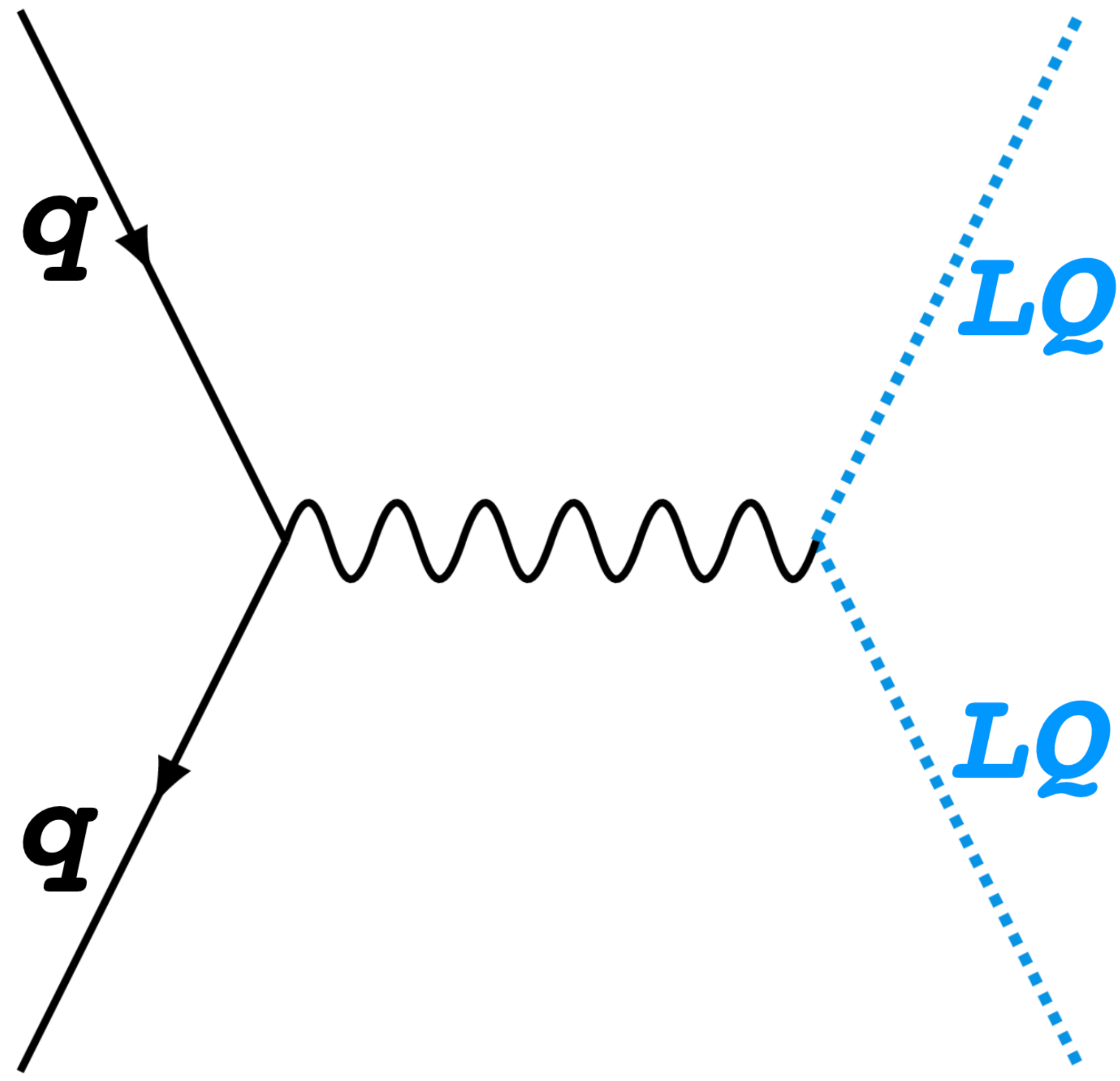} \; \; \; \includegraphics[scale=0.19]{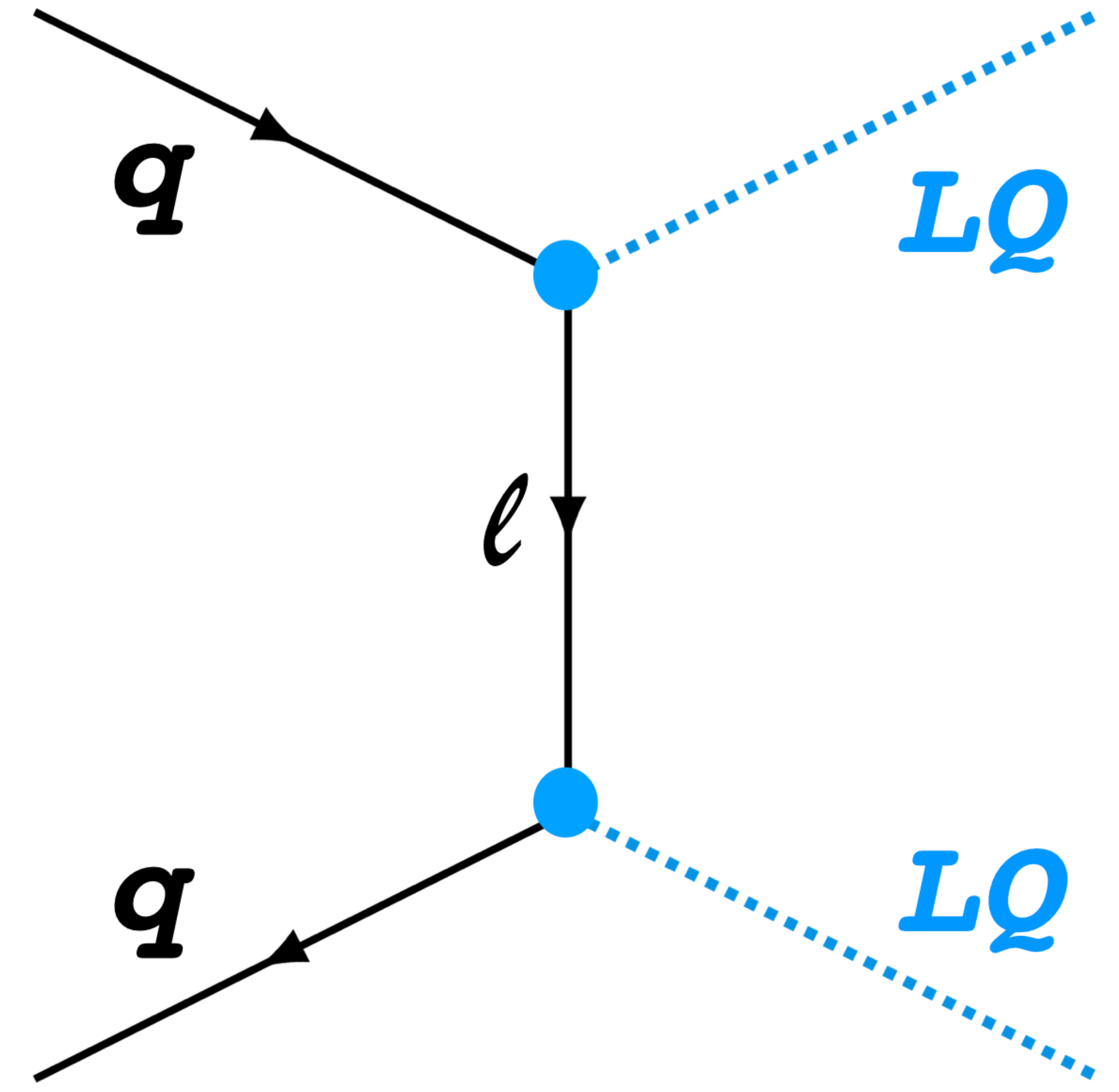} \\
\includegraphics[scale=0.205]{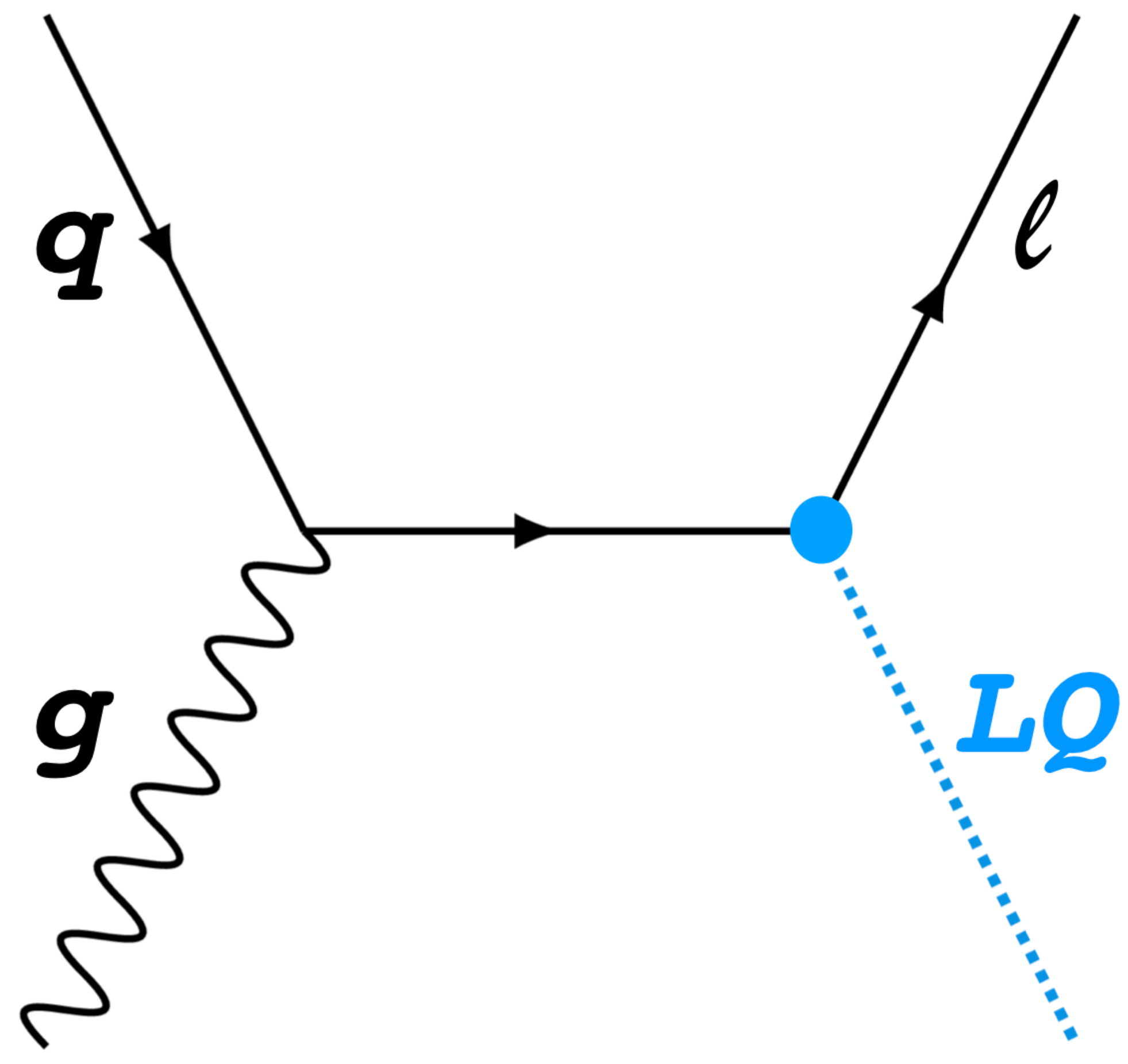} \; \; \; \includegraphics[scale=0.19]{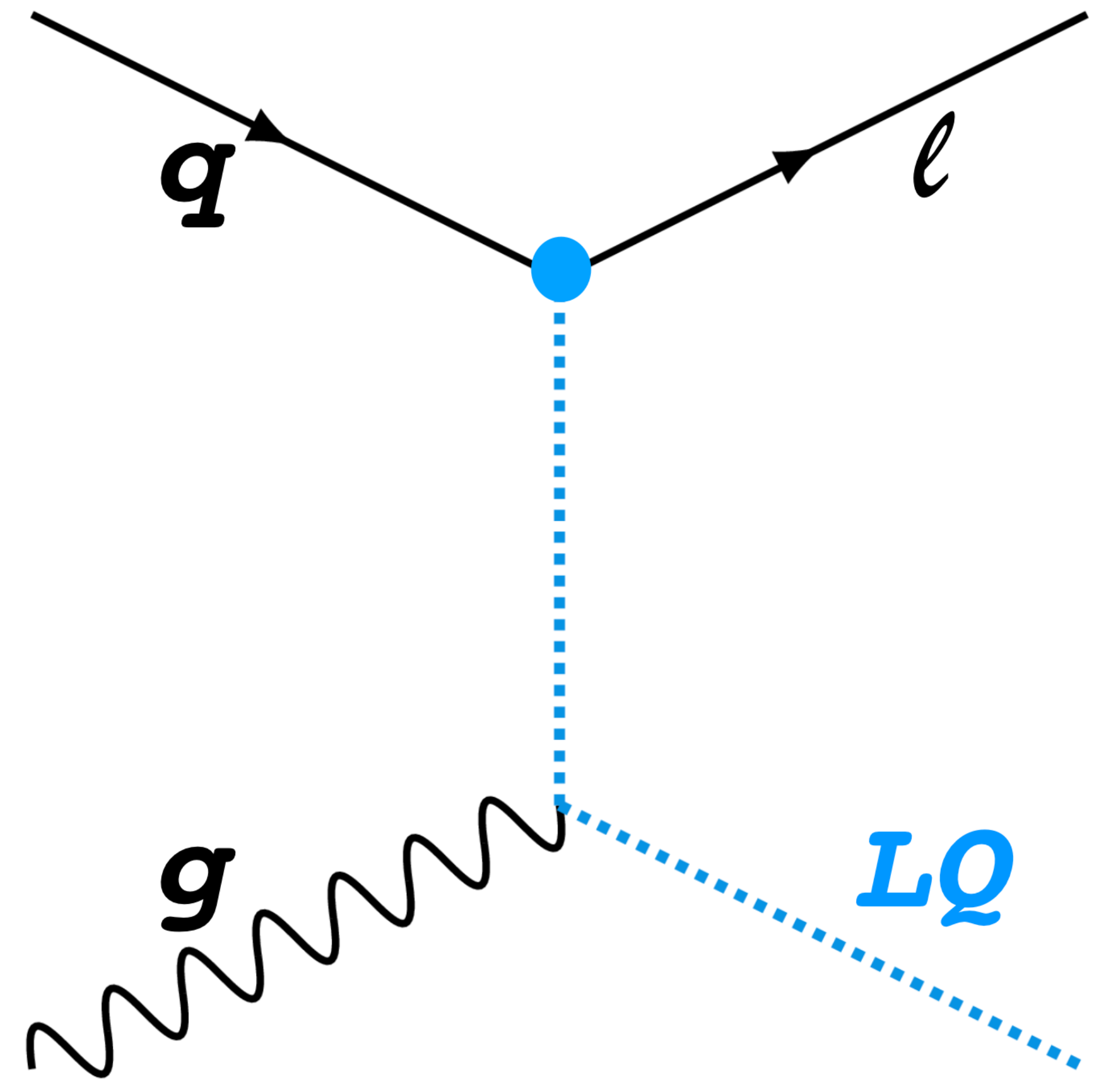} 
\caption{\label{fig:diagram} A sample of leading order Feynman diagrams for the scalar LQ pair production (top row) and single scalar LQ plus lepton production (bottom row).}
\end{figure}

The physics of LQs is a mature subject and its roots date back to the introduction of unification of the quarks and leptons of the SM~\cite{Pati:1973uk}. There exists a number of in-depth reviews of various aspects of the LQ physics one can consult~\cite{Davidson:1993qk,Hewett:1997ce,Nath:2006ut,Dorsner:2016wpm}. These aspects are related to the flavour physics effects, collider physics signatures, and proton decay signals, to name a few. In this note we revisit the production mechanisms of LQs in the proton-proton collisions in view of the need for an up-to-date Monte Carlo event generator output that can be used for the current and future experimental searches and search recasts~\cite{Dorsner:2014axa,Mandal:2015vfa,Diaz:2017lit,Bandyopadhyay:2018syt}. We especially address the single LQ production in association with a lepton and the LQ pair production including important next-to-leading order (NLO) QCD corrections. A sample of leading order (LO) Feynman diagrams for these processes involving scalar LQs is shown in figure~\ref{fig:diagram}.

We stress from the onset that there already exist explicit calculations of the pair production of scalar LQs at the NLO level~\cite{Kramer:2004df,Mandal:2015lca} as well as several studies of the NLO effects on the single LQ production~\cite{Alves:2002tj,Hammett:2015sea,Mandal:2015vfa} at the LHC. One of our aims is to fill in the missing pieces with respect to the latter process, especially in the context of the sea quark initiated production. In fact, it is very important to entertain a possibility of an LQ dominantly coupled to heavy fermions as motivated by the pattern of fermion masses and mixing parameters, and as recently suggested by the hints on lepton flavour universality violation in $B$-meson decays. (See, for example, ref.~\cite{Buttazzo:2017ixm} for more details.) With this possibility in mind we also address single production of vector LQs through the bottom-gluon fusion processes. 

Since the number of LQs is finite one can easily classify them~\cite{Buchmuller:1986zs}. We provide, as an integral part of this analysis, ready-to-use universal \FeynRules (\UFO)~\cite{Alloul:2013bka} model files for all scalar LQs as well as one vector LQ that are particularly suited for the flavour dependent studies of the LQ signatures at colliders within the \MadGraph~\cite{Alwall:2014hca} framework. We validate our numerical results with the existing NLO calculations for the pair production and present novel results for the single production of scalar (vector) LQs at the NLO (LO) level. These results, in our view, can be particularly useful for the current and future LHC data analyses and accurate search recasts. The \UFO model files are publicly available at {\color{blue}http://lqnlo.hepforge.org}.

The outline of the manuscript is as follows. We present the set-up for our LQ signature studies in section~\ref{sec:implementation}. This is followed by section~\ref{sec:analysis} on numerical analysis that is subdivided into the LQ pair production subsection and the single LQ + lepton production subsection. The strategy for LQ searches inferred from $B$-physics anomalies is described in section~\ref{sec:B-anomalies}. We present our conclusions in section~\ref{sec:conclusions}. Most of our numerical results are summarised in appendix~\ref{appendix:A}.



\section{Implementation and validation}
\label{sec:implementation}

\subsection{Scalar LQ set-up}

The scalar LQ models we implement comprise $S_3 \equiv (\overline{\mathbf{3}},\mathbf{3},1/3)$, $R_2 \equiv (\mathbf{3},\mathbf{2},7/6)$, $\tilde{R}_2 \equiv (\mathbf{3},\mathbf{2},1/6)$, $\tilde{S}_1 \equiv (\overline{\mathbf{3}},\mathbf{1},4/3)$, and $S_1 \equiv (\overline{\mathbf{3}},\mathbf{1},1/3)$, where we specify transformation properties under the SM gauge group. First (second) integer in the brackets corresponds to the dimension of the irreducible representation of $SU(3)$ ($SU(2)$) that the LQ belongs to whereas the rational number is the LQ $U(1)$ hypercharge. Our hypercharge normalisation is such that the electric charge of $S_1$ is $1/3$ in units of the absolute value of the electron charge. We assume that lepton number and baryon number are conserved quantities.

We use \FeynRules~2.0~\cite{Alloul:2013bka} to prepare the model files for each LQ representation. The inclusion of NLO QCD corrections is possible in modern Monte Carlo frameworks that are capable of the automated generation of the corresponding born, one-loop and real matrix elements, and subtraction of infrared singularities. To this purpose, we use the \NLOCT package (version 1.02)~\cite{Degrande:2014vpa} together with \FeynArts (version 3.9)~\cite{Hahn:2000kx} to generate the relevant UV and $R_2$ counterterms at one-loop level in QCD. The resulting models are exported in the \UFO format, which can be directly used within \MadGraph framework, where all required one-loop amplitudes are automatically generated with \MadLoop \cite{Hirschi:2011pa} and \Ninja~\cite{Peraro:2014cba,Hirschi:2016mdz}. The corresponding real amplitudes are generated from the underlying \UFO model, while the infrared subtraction of the real contributions is automatically performed \`a la FKS~\cite{Frixione:1995ms} in \MadFKS \cite{Frederix:2009yq}.

The kinetic and mass terms are implemented in the same manner for all the scalar LQs and are given by
\begin{equation}
\mathcal{L}^{\Phi}_\mathrm{kinetic}=\left(\mathcal{D}_{\mu}\Phi\right)^{\dagger}\left(\mathcal{D}^{\mu}\Phi\right)-m_\mathrm{LQ}^{2}\Phi^{\dagger}\Phi,
\label{eq:kinetic}
\end{equation}
where $\mathcal{D}_{\mu}$ is the appropriate covariant derivative and $\Phi=S_3, R_2, \tilde{R}_2, \tilde{S}_1, S_1$. As implied by the second term in eq.~\eqref{eq:kinetic} all the components within the given LQ multiplet $\Phi$, when $\Phi$ transforms non-trivially under $SU(2)$, are assumed to have the same mass. This assumption is driven by the electroweak precision measurements. (See, for example, section~4.2 in ref.~\cite{Dorsner:2016wpm} for more details.) The mass splitting which generates small enough one-loop correction to oblique $Z$-pole observable can be completely neglected in view of the current direct limits on LQ masses. In other words, one expects correlated signal in searches for the same-mass LQ states with different electric charge (by one unit). A combination of such searches can improve the overall sensitivity with respect to the parameter space of the non-trivial $SU(2)$ LQ multiplet(s).

For the flavour dependent part of the lagrangian we closely follow notation of ref.~\cite{Dorsner:2016wpm} and implement the most general form of Yukawa couplings. The fermion content is taken to be purely that of the SM. We explicitly assume that the unitary transformations of the right-chiral fermions are not physical. In our convention the Cabbibo-Kobayashi-Maskawa (CKM) rotations reside in the up-type quark sector whereas the Pontecorvo-Maki-Nakagawa-Sakata (PMNS) rotations originate from the neutrino sector. These rotations provide connection between Yukawa couplings in the case when LQ interacts with the $SU(2)$ doublet(s) of the SM fermions. For example, one set of the $R_2$ Yukawa couplings that features the CKM matrix $V$ is 
\begin{equation}
\mathcal{L}^{R_2}_\mathrm{Yukawa} \supset +y^{LR}_{2\,ij}\bar{e}_{R}^{i} R_{2}^{a\,*}Q_{L}^{j,a} =+(y^{LR}_2 V^\dagger)_{ij} \bar{e}^i_R 
  u^j_L R_2^{5/3\,*} +y^{LR}_{2\,ij} \bar{e}^i_R d^j_L R_2^{2/3\,*} ,
\label{eq:yukawaR2}
\end{equation}
where $y^{LR}_{2}$ is the $3\times3$ matrix in the flavour space, $Q_{L}$ is a left-chiral quark doublet, $e_R$ is a right-chiral charged lepton, $a=1,2$ is an $SU(2)$ index, and $i,j=1,2,3$ are flavour indices. The couplings of $\tilde{R}_2$, on the other hand, feature the PMNS matrix $U$ since the relevant lagrangian reads
\begin{equation}
\mathcal{L}^{\tilde{R}_2}_\mathrm{Yukawa} \supset -\tilde{y}^{RL}_{2\,ij}\bar{d}_{R}^{i}\tilde{R}_{2}^{a}\epsilon^{ab}L_{L}^{j,b} =-\tilde{y}^{RL}_{2\,ij}\bar{d}_{R}^{i}e_{L}^{j}\tilde{R}_{2}^{2/3}+(\tilde{y}^{RL}_2 U)_{ij}\bar{d}_{R}^{i}\nu_{L}^{j}\tilde{R}_{2}^{-1/3} ,
\label{eq:yukawaR2t}
\end{equation}
where $\tilde{y}^{RL}_{2}$ is the $3\times3$ matrix in the flavour space, $L_{L}$ is a left-chiral lepton doublet, $d_R$ is a right-chiral down-type quark, and $\epsilon^{ab}$ is Levi-Civita symbol. The hermitian conjugate parts are omitted from eqs.~\eqref{eq:yukawaR2} and~\eqref{eq:yukawaR2t} for brevity. Note that our convention allows one to completely neglect the PMNS rotations as the neutrino flavour is not relevant for the processes we are interested in. In the actual model file implementations the PMNS matrix is thus set to be an identity matrix whereas the only relevant angle in the CKM matrix is taken to be the Cabbibo one. These assumptions can be modified using the parameter restriction files that are provided with each LQ model. 

One could also entertain a possibility of introducing one or more right-chiral neutrinos thus extending the SM fermion sector. This would allow one to study one additional scalar LQ state --- $\bar{S}_1 \equiv (\overline{\mathbf{3}},\mathbf{1},-2/3)$ --- and to consider additional sets of Yukawa couplings for $\tilde{R}_2$ and $S_1$~\cite{Dorsner:2016wpm}. These three scalar multiplets have the same transformation properties under the SM gauge group as the squarks, where the role of the right-chiral neutrino(s) could be played by neutralino(s). The right-chiral neutrino introduction would, in principle, yield the same phenomenological signatures that one has for those LQs that couple to the left-chiral neutrinos as long as the right-chiral neutrinos are light enough. This fact and the close correspondence between the LQ and squark properties is often used to reinterpret dedicated experimental searches for supersymmetric particles in terms of limits on the allowed LQ parameter space. See, for example, ref.~\cite{Diaz:2017lit} for a recent recast along these lines. Be that as it may, the model files we provide can be modified to incorporate these hypothetical fermionic fields and associated interactions.   

We always consider a scenario where the SM is extended with a single scalar LQ multiplet. From these single LQ model files one can easily generate more complicated scenarios of new physics (NP) when two or more scalar LQs are simultaneously present at the energies relevant for collider physics. Since the LQ electric charge eigenstates coincide with the mass eigenstates in the single LQ extensions we use this property to uniquely denote a given LQ component. For example, $R_2^{5/3}$ ($\tilde{R}_{2}^{-1/3}$) is denoted as \texttt{R2p53} (\texttt{R2tm13}) in model files. The fact that \FeynRules~2.0~\cite{Alloul:2013bka} does not accommodate antifundamental representation of $SU(3)$ has prompted us to implement all the LQs as triplets of colour in model files.

In the \MadGraph model parameter card of a given LQ scenario one can modify the LQ mass $m_\mathrm{LQ}$ and its Yukawa couplings. For example, the 13 element of the Yukawa coupling matrix $\tilde{y}^{RL}_{2}$ from eq.~\eqref{eq:yukawaR2t} is denoted as \texttt{yRL1x3} in the associated model file. For the total decay width of a given LQ we assume that all the quark masses except the top quark can be neglected and provide relevant expressions. Note that mass eigenstates that originate from the same LQ multiplet do not need to have the same decay width. To that end we denote the associated decay widths differently. For example, the decay width of $R_2^{5/3}$ ($R_2^{2/3}$) is denoted as \texttt{WR253} (\texttt{WR223}).

In order to validate the NLO QCD implementation we generate the LQ decay process with \MadGraph. This calculation consists of one born, one virtual and two real-radiation diagrams. The analytic formula for the partial decay width for massless fermions and lagrangian defined as $\mathcal{L} \supset (- y_{q\ell} ~\bar q P_{L,R} \ell ~\Phi ~+$~h.c.) is ~\cite{Plehn:1997az}
\begin{equation}
\label{eq:DecayWidth}
\Gamma(\Phi \to q \ell) = \frac{|y_{q\ell}|^2 m_\mathrm{LQ}}{16 \pi} \left( 1 + \left(\frac{9}{2} -\frac{4 \pi^2}{9}\right) \frac{\alpha_s}{\pi} \right)~,
\end{equation}
where $\alpha_s=g_s^2/(4 \pi)$ is the strong coupling constant, $P_{L}$ and $P_{R}$ are the standard left- and right-chiral projection operators, and $y_{q\ell}$ is the Yukawa coupling strength. Our numerical result for the NLO QCD correction factor ($K_F - 1 \approx 0.0043$), obtained using \MadGraph, agrees perfectly with the analytic formula in eq.~\eqref{eq:DecayWidth}. That is, by reproducing finite one-loop corrections, we have validated the implementation of the corresponding QCD counterterms in the \UFO model(s).

\subsection{Vector LQ set-up}

The phenomenology of vector LQ states is sensitive to their UV origin. The only vector LQ UFO implementation we opt to provide here is the one that involves $U_1\equiv (\mathbf{3},\mathbf{1},2/3)$ field. This vector boson has attracted a lot of attention recently~\cite{Buttazzo:2017ixm,DiLuzio:2017vat,Bordone:2017bld,Barbieri:2017tuq} and its model file can be appropriately modified to represent other vector LQ states, if needed, for flavour dependent studies.

The kinetic and mass terms of $U_1$ are
\begin{equation}
\mathcal{L}^{U_1}_\mathrm{kinetic}= -\frac{1}{2} U_{\mu\nu}^{\dagger} U_{\mu\nu} - i g_s \kappa~ U_{1\mu}^{\dagger} T^a U_{1 \nu} G^a_{\mu \nu}+ m_{U_1}^2 U_{1\mu}^\dagger U_{1\mu},
\label{eq:kin}
\end{equation}
where $U_{\mu\nu} = D_\mu  U_{1\nu} - D_\nu  U_{1\mu}$ is a field strength tensor and $\kappa$ is a dimensionless coupling that depends on the UV origin of the vector. For the Yang-Mills case $\kappa =1$, while for the minimal coupling case $\kappa = 0$. Precision calculations with vector LQ require UV completion and this ambiguity is only in part captured by the $\kappa$ dependence that we study in section~\ref{sec:vLQnum}. In fact, unitarization of the high-$p_T$ scattering amplitudes requires additional dynamics not far beyond the LQ mass scale, which can impact the production processes in a non-trivial way. For example, an extra colour octet vector might exist in a complete model and give an $s$-channel contribution to the LQ pair production. Therefore, in this paper, we concentrate on the LO effects in QCD. Implementation of the benchmark UV realisations together with the NLO QCD corrections is left for future work. For a complementary study of NLO QCD effects in vector LQ processes see ref.~\cite{Hammett:2015sea}.

The Yukawa part of the $U_1$ lagrangian is defined in eq.~\eqref{eq:U1} in section~\ref{sec:B-anomalies}. For simplicity, here we implement the following lagrangian $\mathcal{L} \supset (g_{b_L} \bar b_L \gamma^\mu \tau_L U_{1\mu} +  g_{t_L} \bar t_L \gamma^\mu \nu^\tau_L U_{1\mu} + \textrm{h.c.})$ where $g_{b_L}$ ($g_{t_L}$) is the coupling strength of $U_1$ with the bottom-tau (top-neutrino) pair. ($SU(2)$ gauge invariance predicts $g_{b_L}=g_{t_L}$.) The model files can easily be modified to include other interactions. The relevant parameters that one can vary are $m_{U_1}$ (\texttt{mLQ}), $\kappa$ (\texttt{kappa}), $g_{t_L}$ (\texttt{gtL}), and 
$g_{b_L}$ (\texttt{gbL}). The $U_1$ particle (antiparticle) name in the model file is \texttt{vlq} (\texttt{vlq$\sim$}). LQ total decay width is denoted with \texttt{wLQ}, and should be correspondingly adjusted for a given set of input parameters. For example, the LO partial decay width $\Gamma(U_1 \to b \tau^+) = |g_{b_L}|^2 m_{U_1} /(24 \pi)$ if one neglects $b$ and $\tau$ masses. Numerical results using this implementation are presented in section~\ref{sec:vLQnum}.

\section{Numerical analysis}
\label{sec:analysis}

The \UFO implementation at the NLO in QCD allows us to calculate the total inclusive cross section for either LQ pair production or single LQ production within the \MadGraph framework for a given LQ mass. We have also prepared a simplified scalar LQ model file, {named \texttt{Leptokvark\_NLO}}, that can be used to efficiently determine inclusive cross sections at the tree level and NLO level. This simplified model uses the fact that the pair production of any scalar LQ is, for all practical purposes, solely QCD driven whereas the single scalar LQ production in association with lepton depends only on the particular quark flavour that the LQ couples to and the associated coupling strength, as discussed below.


\begin{figure}[tbp]
\centering
\includegraphics[scale=0.355]{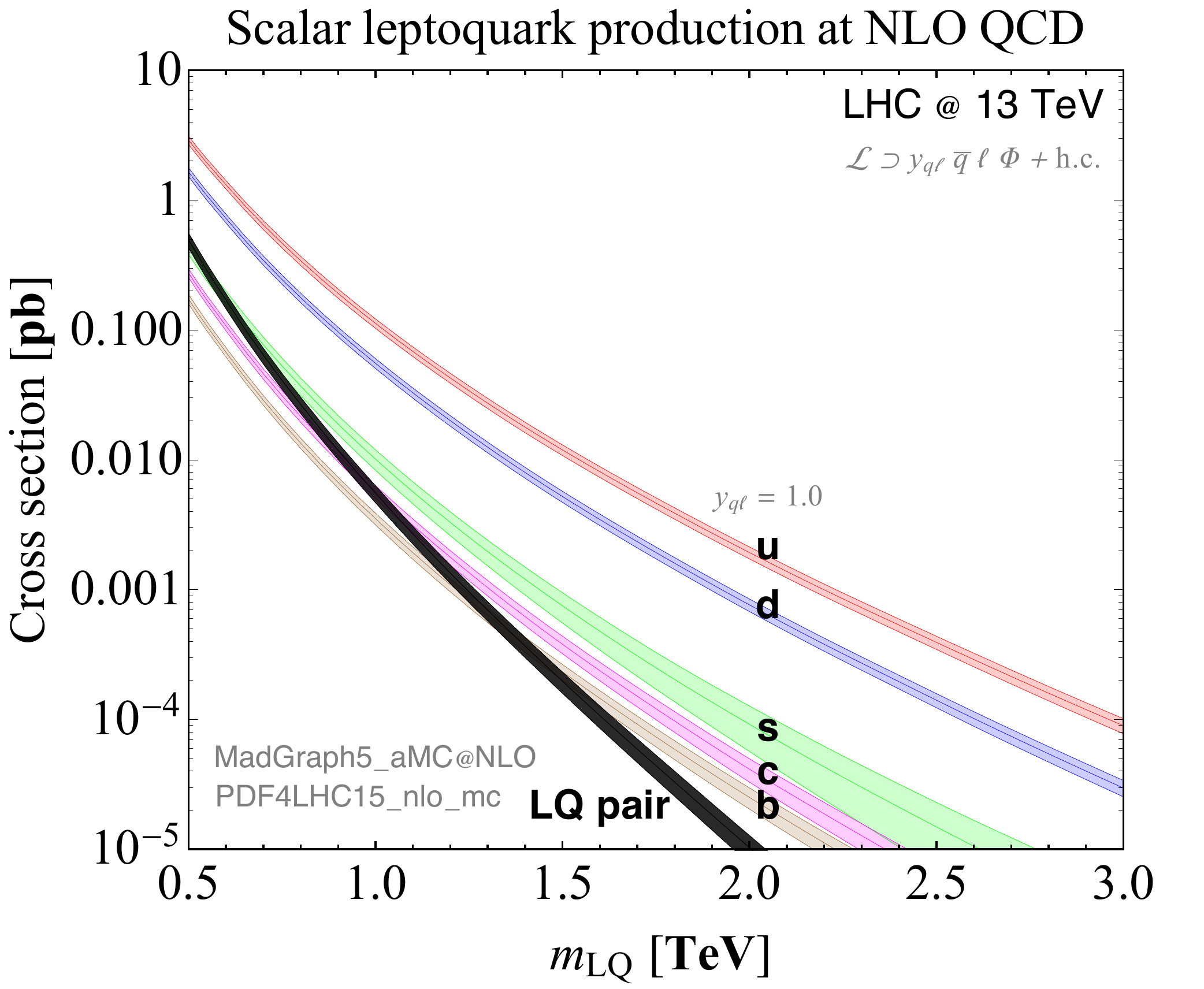}
\includegraphics[scale=0.35]{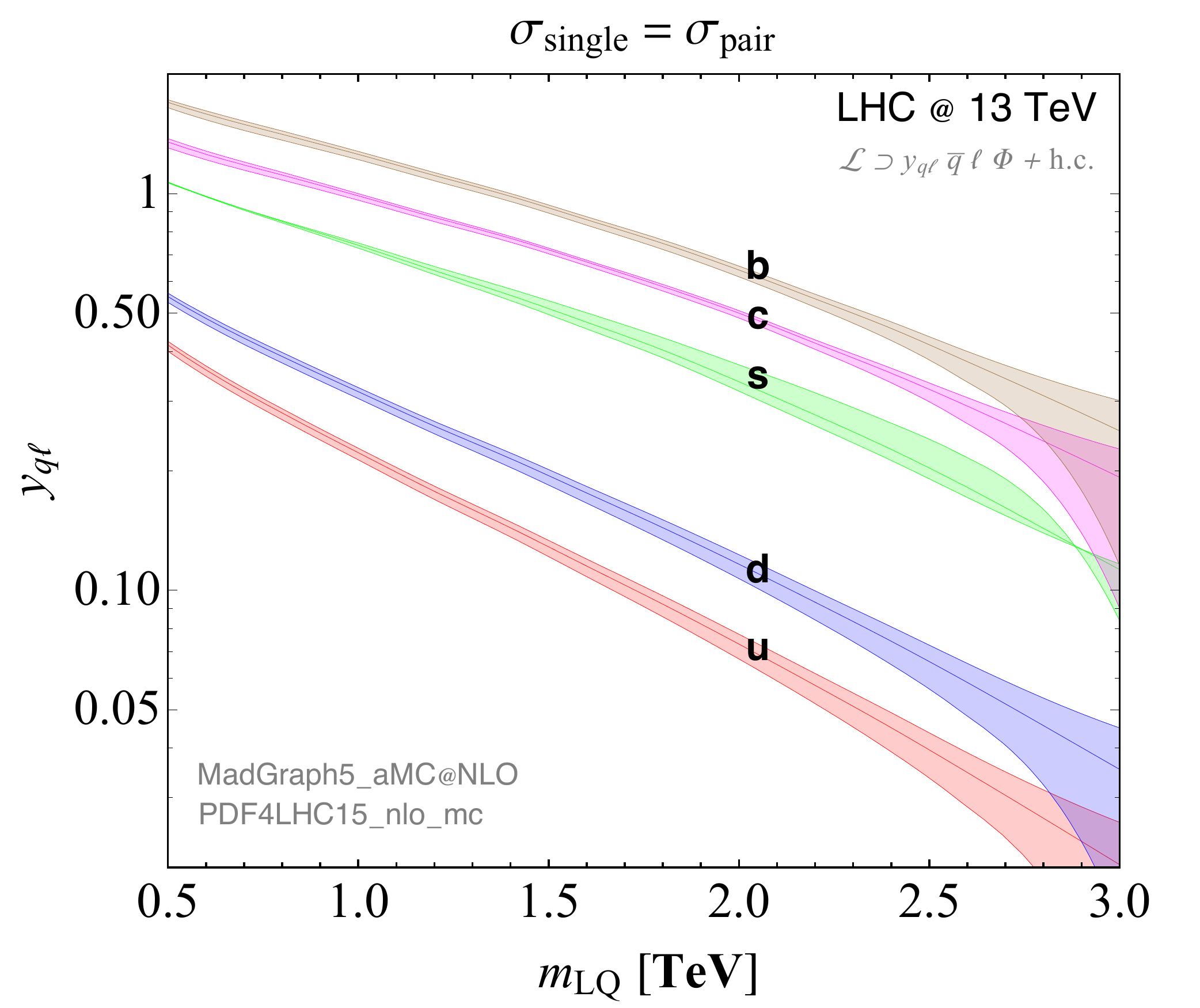}
\caption{\label{fig:all} (Left panel) Total inclusive cross sections (in pb) at NLO in QCD for scalar LQ production in proton-proton collisions using the {\tt PDF4LHC15\_nlo\_mc}~\cite{Butterworth:2015oua} sets at 13\,TeV center-of-mass energy as a function of the LQ mass $m_\mathrm{LQ}$. The central values are obtained using fixed factorisation and renormalisation scales $\mu_F=\mu_R= m_\mathrm{LQ}$. The total uncertainty (shown with bands) is obtained by adding the PDF and perturbative uncertainties in quadrature, where the former one is given by the $68\%$ C.L.\ ranges when averaging over the PDF replicas while the latter one is estimated varying factorisation and renormalisation scales within $\mu_F, \mu_R \in [0.5, 2]~ m_\mathrm{LQ}$. Prediction for the single LQ production ($p p \to \Phi \ell$ plus $p p \to \bar \Phi \bar \ell$) initiated from up, down, strange, charm, and bottom initial-state flavours is marked with {\bf u}, {\bf d}, {\bf s}, {\bf c}, and {\bf b}, respectively, while the LQ pair production ($p p \to \Phi \bar \Phi$) is denoted with {\bf LQ pair}. All single LQ production cross sections correspond to the case when the coupling strength of the LQ to the quark-lepton pair is set to one ($y_{q\ell} = 1$).
(Right panel) $y_{q\ell} = \sqrt{\sigma_{\textrm{pair}}/\sigma_{\textrm{single}} (y_{q\ell}=1})$ as a function of the LQ mass for all initial-state quarks at 13\,TeV. The three lines for each quark flavour are obtained using central, plus, and minus predictions for the total cross sections, and the shaded area indicates the size of prediction uncertainty. We have checked that the contribution of the Yukawa dependent diagram with $t$-channel lepton to LQ pair production is negligible in determining these lines.}
\end{figure}

\subsection{Scalar LQ pair production}
\label{sec:pair}

LQ pair production at hadron collider(s) is a QCD driven process that is, at this point, completely determined by the LQ mass and strong coupling constant due to the existing experimental measurements on atomic parity non-conservation~\cite{Guena:2004sq,Wood:1997zq} and the current direct search limits on LQ masses at the LHC. The atomic parity non-conservation measurements limit the allowed strength of the LQ couplings to the first generation of quarks and leptons~\cite{Dorsner:2016wpm,Dorsner:2014axa}. These need to be small and, as such, cannot affect LQ pair production. One might think that it could be possible to affect pair production with the large enough Yukawa couplings to the second and/or third generation of quarks thereby avoiding atomic parity violation constraints. Another option is to have couplings between the first generation quarks and second and/or third generation leptons. These particular possibilities are of limited interest due to the fact that for such large couplings (and masses) single LQ process is expected to be dominant. (See, for example, figure~3 in ref.~\cite{Dorsner:2014axa}.) We thus completely neglect $t$-channel contribution towards pair production of LQs in our numerical studies. The Feynman diagram that depicts the contribution that we neglect is shown in the third panel of the first row of figure~\ref{fig:diagram}.

The dominant pair production mechanism at the LHC is a gluon-gluon scattering followed by the quark-antiquark annihilation with the representative Feynman diagrams shown in the first and second panel of the first row of figure~\ref{fig:diagram}, respectively. The latter process grows in importance as the LQ mass increases. Differential and integral cross sections for these processes at the tree level~\cite{Blumlein:1996qp} and NLO level~\cite{Kramer:2004df} are well-known. We find perfect agreement between our results and analytic expressions that are available in the literature for the same choice of PDFs.  

We use our simplified model file to evaluate total inclusive cross section $\sigma^\mathrm{pair}$ at the NLO level as a function of LQ mass $m_\mathrm{LQ}$ for 13\,TeV, 14\,TeV, and 27\,TeV center-of-mass energies for the proton-proton collisions for the {\tt PDF4LHC15} PDF sets~\cite{Butterworth:2015oua} using \MadGraph. Again, all the model files we provide for scalar LQs yield exactly the same result. The cross section dependancy on the renormalisation and factorisation scale variations is also taken into account in our evaluation, as well as uncertainty due to the PDF determination. The following scale variations are used in this determination, $\mu_R,\mu_F=m_\mathrm{LQ}/2, m_\mathrm{LQ}, 2 m_\mathrm{LQ}$, using the method of ref.~\cite{Alwall:2014hca}. The relevant NLO results for $\sigma^\mathrm{pair}$ are summarised in appendix~\ref{appendix:A} in tables~\ref{tab:13PDF4LHC15},~\ref{tab:14PDF4LHC15}, and~\ref{tab:27PDF4LHC15} for 13\,TeV, 14\,TeV, and 27\,TeV, respectively, where we also present the PDF uncertainty. The uncertainties are quoted in per cent units with respect to the cross section central value. 

Finally, we present in figure~\ref{fig:all} total inclusive cross section at NLO in QCD for scalar LQ pair production (black band) in proton-proton collisions at 13\,TeV center-of-mass energy as a function of the LQ mass $m_\mathrm{LQ}$ for the {\tt PDF4LHC15\_nlo\_mc}~\cite{Butterworth:2015oua} PDF sets.

\begin{figure}[tbp]
\centering
\includegraphics[scale=0.6]{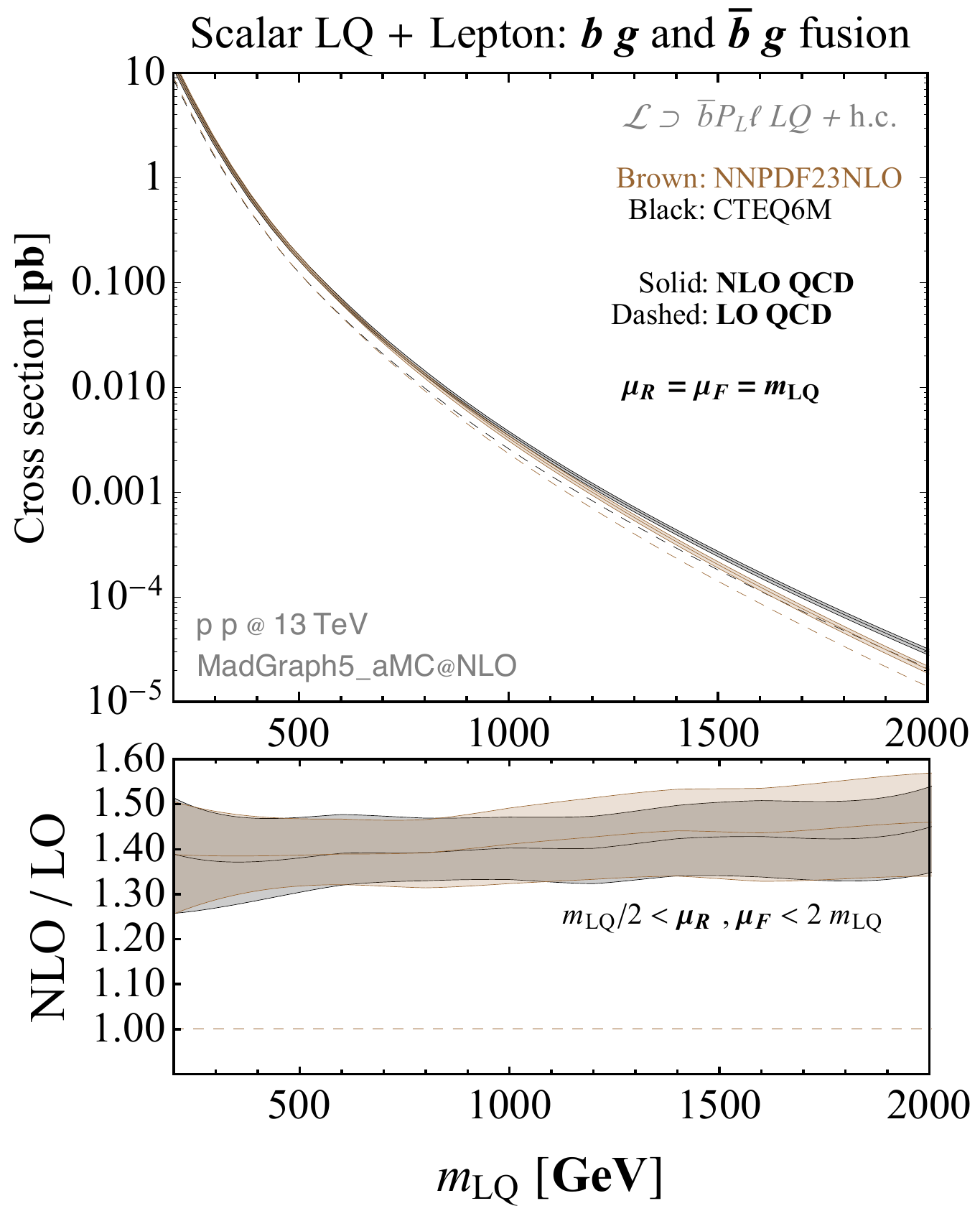}
\caption{\label{fig:single} Total inclusive single LQ + lepton production cross sections (from bottom flavour) at the NLO in QCD (upper panel) and ratio of NLO and LO cross sections (lower panel) for the proton-proton collisions for the {\tt NNPDF23NLO}~\cite{Ball:2014uwa} and {\tt CTEQ6M}~\cite{Pumplin:2002vw} sets at 13\,TeV center-of-mass energy as a function of the LQ mass $m_\mathrm{LQ}$. The LQ coupling strength to bottom quark and lepton is set to one ($y_{b \ell} = 1$).}
\end{figure}

\subsection{Single scalar LQ production}
\label{sec:single}

The single LQ production in association with a lepton at tree level is induced via partonic process $g q \to \ell \Phi$ and involves two diagrams that are shown in the second row of figure~\ref{fig:diagram}. The NLO QCD corrections to this process involve virtual one-loop and real-radiation diagrams, which all have the same linear dependence on the Yukawa coupling $y_{q\ell}$ defined through $\mathcal{L} \supset ( y_{q\ell} ~\bar q P_{L,R} \ell ~\Phi ~+$~h.c.). The interference effects that might be relevant when LQ simultaneously couples to two fermion pairs of the same flavour but different chirality structure are suppressed by the final-state lepton mass and can thus be safely neglected. Therefore, the inclusive NLO QCD K-factor is rather model independent, i.e., it does not depend on the specific LQ representation, nor the final-state lepton flavour, chirality, and charge. It only depends on the flavour of the initial-state quark in the tree-level diagram, the LQ mass $m_\mathrm{LQ}$, and trivially on the coupling since $\sigma^\mathrm{single} \sim |y_{q\ell}|^2$.

We use our simplified NLO model file to evaluate total inclusive single LQ production cross sections $\sigma^\mathrm{single}_{u,d,s,c,b}$ for the proton-proton collisions using the \PDFLHC PDF sets~\cite{Butterworth:2015oua}. These cross sections are due to production through the corresponding quark flavour, as indicated in the subscript, where we set the associated Yukawa coupling strength to one ($y_{q\ell} = 1$), and vary the LQ mass $m_\mathrm{LQ}$. Again, the single LQ production cross section is proportional to a square of the coupling strength and can thus be trivially rescaled. We furthermore evaluate $\sigma^\mathrm{single}_{u,d,s,c,b}$ for 13\,TeV, 14\,TeV, and 27\,TeV center-of-mass energies. The cross section dependancy on the renormalisation and factorisation scale variations is also taken into account in this evaluation following the method of ref.~\cite{Alwall:2014hca}. The relevant NLO results are summarised in appendix~\ref{appendix:A} in tables~\ref{tab:13PDF4LHC15},~\ref{tab:14PDF4LHC15}, and~\ref{tab:27PDF4LHC15} for 13\,TeV, 14\,TeV, and 27\,TeV, respectively, where we also present the PDF uncertainty. The uncertainties are quoted in per cent units with respect to the cross section central value.

We present total inclusive single LQ production cross sections in the left panel of figure~\ref{fig:all} in proton-proton collisions at 13\,TeV center-of-mass energy as a function of the LQ mass $m_\mathrm{LQ}$ for the {\tt PDF4LHC15\_nlo\_mc}~\cite{Butterworth:2015oua} PDF sets. We, again, take that the relevant Yukawa coupling strength between a leptoquark and a quark-lepton pair equal to one ($y_{q\ell} = 1$). In the right panel of figure~\ref{fig:all} we show what values of Yukawa couplings one needs to use to have equality between the total inclusive single LQ production cross section and the total inclusive LQ pair production cross section for a given initial quark flavour as a function of the LQ mass. This plot clearly shows the importance of single LQ production in the heavy LQ regime.

We furthermore present in figure~\ref{fig:single} total inclusive single LQ production cross sections at the NLO level (upper panel) and a ratio of NLO and LO cross sections (lower panel) for the {\tt NNPDF23NLO}~\cite{Ball:2014uwa} and {\tt CTEQ6M}~\cite{Pumplin:2002vw} sets at 13\,TeV proton-proton center-of-mass energy as a function of the LQ mass $m_\mathrm{LQ}$. The lower panel of figure~\ref{fig:single} shows the K-factor for these PDF sets.

\subsection{Single vector LQ production from $b$ quark}
\label{sec:vLQnum}

\begin{figure}[tbp]
\centering
\includegraphics[scale=1]{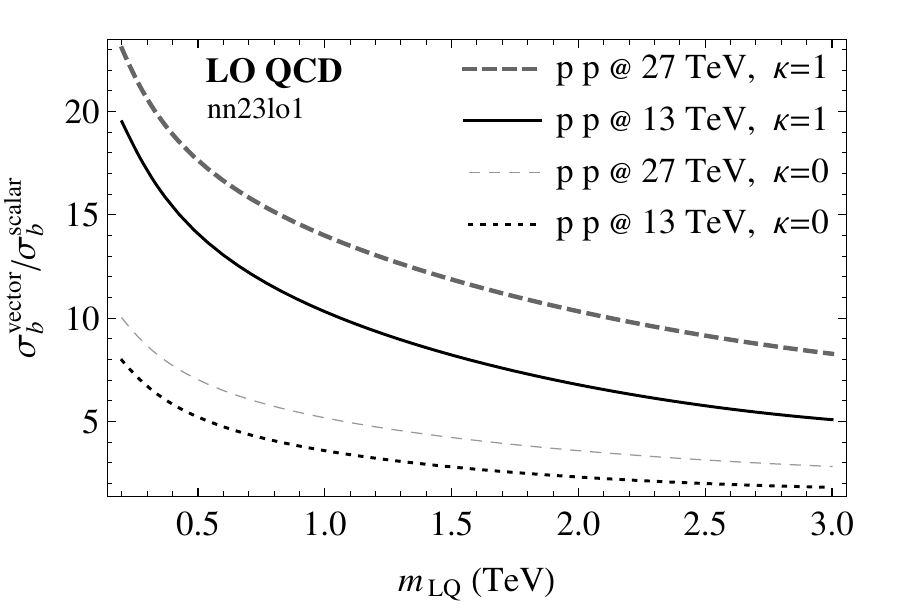}
\caption{\label{fig:vector} Ratio of total inclusive cross sections at the LO in QCD for the single production of a vector LQ ($\sigma_b^\mathrm{vector}$) and a scalar LQ ($\sigma_b^\mathrm{scalar}$) through a fusion of $b$ and $\bar{b}$ quarks with gluons for {\tt nn23lo1} PDFs in proton-proton collisions at 13\,TeV and 27\,TeV as a function of the LQ mass $m_\mathrm{LQ}$. We present $\sigma_b^\mathrm{vector}/\sigma_b^\mathrm{scalar}$ for both the Yang-Mills case $\kappa =1$ and the minimal coupling case $\kappa = 0$.}
\end{figure}

In this numerical exercise we study the dependence of the single vector LQ plus lepton production on the (adjustable) non-minimal QCD coupling $\kappa$ defined in eq.~\eqref{eq:kin}.
We present in figure~\ref{fig:vector} ratio of total inclusive cross sections at the LO in QCD for the single production of a vector LQ and a scalar LQ through a fusion of $b$ and $\bar{b}$ quarks with gluons in proton-proton collisions at 13\,TeV and 27\,TeV as a function of the LQ mass $m_\mathrm{LQ}$. We explicitly take that both scalar and vector LQs couple to the bottom-tau pair with the same Yukawa coupling strength ($g_{b_{L}}=y_{b \tau}$). Since $\sigma_b^\mathrm{vector}$ and $\sigma_b^\mathrm{scalar}$ scale in the same way with regard to the Yukawa coupling the ratio $\sigma_b^\mathrm{vector}/\sigma_b^\mathrm{scalar}$ we present in figure~\ref{fig:vector} is Yukawa coupling independent. This simply means that the knowledge of the total inclusive cross section for the single production of a scalar LQ, for a given strength of Yukawa coupling, allows one to obtain corresponding cross section for vector LQ. We consider both the Yang-Mills case $\kappa =1$ and the minimal coupling case $\kappa = 0$ to capture $\kappa$ dependance. We find that for a fixed $m_\mathrm{LQ}$ the ratio $\sigma_b^\mathrm{vector}/\sigma_b^\mathrm{scalar}$ grows with the increase in value of $\kappa$ parameter. Note that the ratio $\sigma_b^\mathrm{vector}/\sigma_b^\mathrm{scalar}$ is parton distribution function (PDF) insensitive and its value decreases as the LQ mass increases. We, for definiteness, use {\tt nn23lo1} PDFs to perform the numerical calculation. The cross sections are evaluated for $\mu_R,\mu_F=m_\mathrm{LQ}$, where $\mu_R$ ($\mu_F$) is the renormalisation (factorisation) scale. 

\section{$B$-anomalies inspired LQ search strategy}
\label{sec:B-anomalies}

Semileptonic $B$-meson decays have recently received a lot of attention in view of an increasing set of experimental measurements that contradict the SM predictions. Despite the fact that a convincing evidence for NP is still missing, the case for it looks very promising as the coherent picture of deviations seems to be solidifying. (See, for example, ref.~\cite{Buttazzo:2017ixm} for more details.) While the experimental and theoretical endeavour in $B$-physics slowly keeps moving forward, it is important and timely to provide consistent NP scenarios or, better still, NP models that are able to predict smoking gun signatures in other (ongoing) searches, in particular, at the high-$p_T$ frontier experiments, such as ATLAS and CMS. 

Anomalies in $B$-meson decays consistently point to a violation of lepton flavour universality (LFU) and can be grouped into two different classes. These are (i) deviations from $\tau / \ell$ (where $\ell= e,\mu$) universality in semi-tauonic decays as defined by $R(D^{(*)})$ observables ($b \to c \ell \nu$ charged currents)~\cite{Lees:2013uzd,Hirose:2016wfn,Aaij:2015yra} and (ii) deviations from $\mu / e$ universality in rare decays as defined by $R(K^{(*)})$ observables ($b \to s \ell \ell$ neutral currents)~\cite{Aaij:2014ora,Aaij:2017vbb}. Further evidence of coherent deviations in rare $b \to s \mu \mu$ transitions has been observed in the measurements of angular distributions of $B \to K^* \mu^+ \mu^-$~\cite{Aaij:2013qta,Aaij:2015oid}. The overall statistical significance of the discrepancies in the clean LFU observables alone is at the level of $4\,\sigma$ for both charged and neutral current processes. See, for example, refs.~\cite{Capdevila:2017bsm,Altmannshofer:2017yso,Geng:2017svp,Amhis:2016xyh,DAmico:2017mtc}.

\begin{figure}[tbp]
\centering
\includegraphics[scale=0.6]{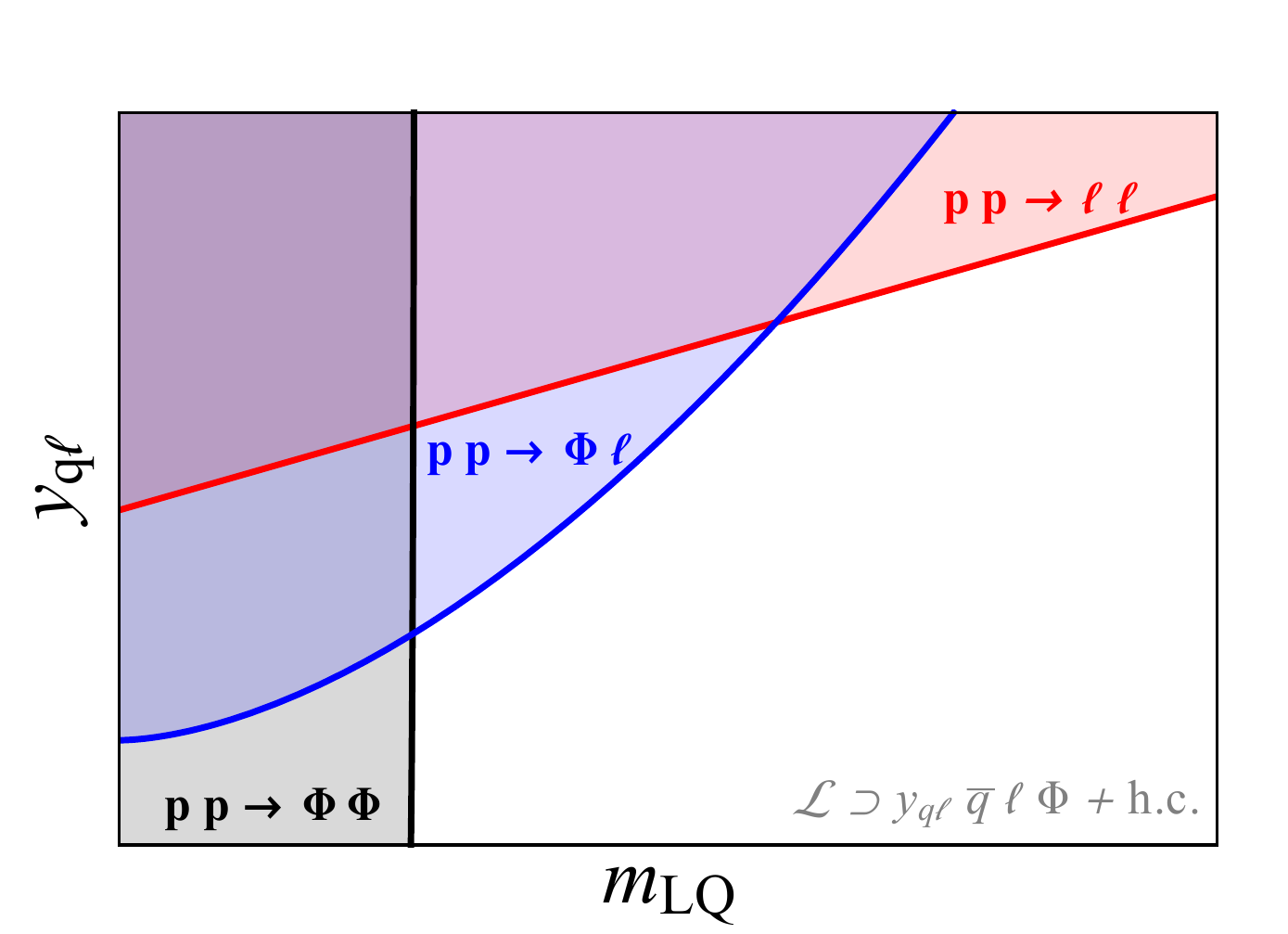}
\caption{\label{fig:strategy} Complementarity illustration for the three LQ processes at the LHC on the $(m_{\rm{LQ}},y_{q \ell})$ parameter space. LQ pair production, dominated by QCD, is (largely) insensitive on the coupling $y_{q \ell}$, setting therefore a robust lower limit on the LQ mass $m_{\rm{LQ}}$. At the opposite end of the LQ mass spectrum, the strongest bound comes from the Drell-Yan production of a dilepton pair via a $t-$channel LQ exchange since amplitude scales with $y_{q \ell}^2$. Finally, in the intermediate mass range, production of a single LQ in association with a lepton is expected to be the most sensitive probe as the associated amplitude scales linearly with $y_{q \ell}$.}
\end{figure}

\textbf{$R(D^{(*)})$ anomaly}: The enhancement of $\mathcal{O}(20\%)$ on top of the SM tree-level CKM-favoured contribution to $b \to c \tau \nu$ transition requires large NP effect that is, presumably, tree-level generated. A careful consideration based on the perturbative unitarity implies that the scale of NP is rather low~\cite{DiLuzio:2017chi}, i.e., in the TeV ballpark, making it an ideal physics case for the LHC. Analysis of the low-energy process in the SM effective field theory (SM EFT) requires NP in (at least) one of the $d=6$ semileptonic four-fermion operators $\mathcal{O}_{V_L}\sim (\bar Q_L \gamma_\mu \sigma^k Q_L) (\bar L_L \gamma^\mu \sigma^k L_L)$, $\mathcal{O}_{S_R}\sim (\bar d_R Q_L) (\bar L_L e_R)$, $\mathcal{O}_{S_L}\sim (\bar Q_L u_R) i \sigma^2 (\bar L_L e_R)$, and $\mathcal{O}_{T}\sim (\bar Q_L \sigma_{\mu \nu} u_R) i \sigma^2 (\bar L_L \sigma^{\mu \nu} e_R)$, where $\sigma^k$, $k=1,2,3$, are Pauli matrices, and $u_R$ are the right-chiral up-type quark fields. For example, a very good fit to data is obtained with a shift in $\mathcal{O}_{V_L}$ operator only, giving a universal enhancement in all $b \to c \tau \nu$ processes. Nonetheless, several other scenarios are fitting data well~\cite{Freytsis:2015qca}. 

In the simplest case, these effective operators can be generated by a tree-level exchange of a single mediator, defining simplified benchmark models for the LHC studies. These include triplet vector (scalar) $U_3^\mu \equiv ({\bf 3},{\bf 3} ,2/3)$ ($S_3 \equiv (\bar {\bf 3},{\bf 3},1/3)$), doublet vector (scalar) $V_2^\mu \equiv ({\bf \bar 3},{\bf 2},5/6)$ ($R_2 \equiv ({\bf 3},{\bf 2},7/6)$), and singlet vector (scalar) $U_1^\mu \equiv ({\bf 3},{\bf 1},2/3)$ ($S_1 \equiv (\bar {\bf 3},{\bf 1},1/3)$)~\cite{Dorsner:2016wpm}. Triplets induce $\mathcal{O}_{V_L}$ operator only, when integrated out, while, for example, $U_1^\mu$ induces $\mathcal{O}_{V_L}$ and $\mathcal{O}_{S_R}$, and $S_1$ yields $\mathcal{O}_{V_L}$ and $\mathcal{O}_{S_L}-1/4~\mathcal{O}_{T}$.

Since the scale required to fit $R(D^{(*)})$ is rather low the main challenge is not only to reconcile it with the non-observation of other related signals such as flavour changing neutral currents (FCNC) in the down-type quark sector (e.g.~\cite{Alonso:2015sja,Greljo:2015mma,Calibbi:2015kma,Crivellin:2017zlb}) or other tree-level flavour changing processes (e.g.~\cite{Alonso:2016oyd}) but also with $\tau$ decays and electroweak precision observables~\cite{Feruglio:2016gvd,Feruglio:2017rjo} as well as high-$p_T$ production of $\tau$ leptons~\cite{Faroughy:2016osc}. These constraints suggest that the LQ couples dominantly (but not entirely) to the third generation fermions. A typical coupling fitting the anomaly is given by $y_{b \tau} \approx m_\mathrm{LQ}/$1\,TeV.

There are two important implications of this discussion. First, the value of the $\Phi$--$q$--$\ell$ coupling $y_{q\ell}$ emerging from the low-energy fit suggests that three types of LQ processes are relevant at the LHC. In addition to the widely studied LQ pair production, single LQ plus $\tau$ lepton production (from initial $b$ quark) turns out to be crucial as indicated in the right panel of figure~\ref{fig:all}. Moreover, as shown in ref.~\cite{Faroughy:2016osc}, a virtual LQ exchange in $t$-channel can give a sizeable contribution to $\tau^+ \tau^-$ lepton pair production in the limit of complete alignment with the down-type quarks. These three processes scale differently with the coupling $y_{q\ell}$ and can thus provide complementary information. Possible exclusion plot one could potentially generate by using this feature is sketched in figure~\ref{fig:strategy}. It is, in our view, crucial to perform such a combined analysis to scrutinise the available parameter space as much as possible.

Second, the decay channels of the LQ resonances are predicted. Let us illustrate this point on a few examples. We, in particular, consider $U_1$, $S_1$, and $S_3$, with the corresponding interactions
\begin{align}
\label{eq:main_S_3}
\mathcal{L}_{S_3} \supset & ~y^{LL}_{3\,ij}\bar{Q}_{L}^{C\,i,a} \epsilon^{ab} (\sigma^k S^k_{3})^{bc} L_{L}^{j,c}+\textrm{h.c.}~,\\
\mathcal{L}_{S_1}  \supset &~y^{LL}_{1\,ij}\bar{Q}_{L}^{C\,i,a} S_{1} \epsilon^{ab}L_{L}^{j,b}+y^{RR}_{1\,ij}\bar{u}_{R}^{C\,i} S_{1} e_{R}^{j}+\textrm{h.c.}~,\\
\mathcal{L}_{U_1} \supset& ~x^{LL}_{1\,ij}\bar{Q}_{L}^{i,a} \gamma^\mu U_{1\mu} L_{L}^{j,a} + x^{RR}_{1\,ij}\bar{d}_{R}^{i} \gamma^\mu U_{1\mu} e_{R}^{j}+\textrm{h.c.}~. \label{eq:U1}
\end{align}
For instance, $U_1^\mu$ decays dominantly to $b \tau$ and $t \nu$ final states. If no right-chiral couplings ($x^{RR}_{1}$) are present, the branching ratios are predicted to be $\mathcal{B}(U_1\to b \tau) = \mathcal{B}(U_1 \to t \nu) = 0.5$\,, motivating a search for $t \nu b \tau$ final state in addition to conventional $b b \tau \tau$ and $t t \nu \nu$ searches~\cite{Aad:2015caa,Sirunyan:2017yrk}. This changes with the inclusion of sizeable $x^{RR}_{1\,3 3}$ in favour of $U_1 \to b \tau$ decay. On the contrary, $S_1$ resonance decays to $b \nu$ and $t \tau$ final states. As in the previous example, the exact branching ratios depend on the relative strengths of the left- and right- couplings. Another very instructive example is that of the scalar triplet $S_3$, which has three degenerate resonances of a different charge. The decay modes and branching ratios are fixed assuming the dominant coupling to be $y^{LL}_{3\,33}$, in particular, $\mathcal{B}(S_3^{1/3}\to b \nu) = \mathcal{B}(S_3^{1/3} \to t \tau) = 0.5$\,, while $\mathcal{B}(S_3^{2/3}\to t \nu) = 1.0$\,, and $\mathcal{B}(S_3^{4/3}\to b \tau) = 1.0$\,. As illustrated by these examples, the high-p$_T$ searches might also prove useful to distinguish the underlying LQ model.

 {The above discussion implicitly assumed the dominant LQ couplings to be with the third family, as predicted in most models with conventional flavour structure, and as (usually) required by the FCNC constraints. However, a viable possibility in some models is to have sizeable coupling to $q_2$--$\ell_3$ fermion current. Here, the LQ tends to decay to light jets (as opposed to $b$ and $t$) and the production from initial $s$ ($c$) flavours (as opposed to $b$) is preferred. Such example has been studied in ref.~\cite{Dorsner:2017ufx} utilising large $y_{s \tau}$ coupling.}

 \textbf{$R(K^{(*)})$ anomaly}:
Rare $B$-decays are generated at one-loop level in the SM, and suffer from additional CKM and GIM suppression. Therefore, the effective scale indicated by the anomaly in rare $b \to s \ell \ell$ transitions is $m_{\Phi}/\sqrt{y_{s \mu} y_{b \mu}} \sim 30$\,TeV, where $y_{q \mu}$ is the relevant $\Phi$--$q$--$\ell$ coupling. If the anomalies are in muons, as suggested by the angular observables, then $V-A$ operator structure has to be generated. At tree level, this can be achieved by an exchange of a single mediator such as $S_3$, $U_1^\mu$ or $U_3^\mu$.

The main implication of such a large effective NP scale is that the LQ pair production might be the only relevant process at the LHC (unless one of the two couplings, i.e., $y_{s \mu}$ or $y_{b \mu}$, is extremely large, or the LQ couples to the valance quarks~\cite{Greljo:2017vvb}). This, however, is not the case at the future circular hadron collider (FCC-hh), where a much heavier LQ could potentially be probed. (See figure~10 of ref.~\cite{Allanach:2017bta}.) On the other hand, the width of an LQ at the TeV scale could be dominated by other decay channels (other than $b \mu$ or $s \mu$). For instance, the interesting option is the decay to third family. Indeed given the same effective $V-A$ structure, a coherent picture of $B$-anomalies is emerging~\cite{Buttazzo:2017ixm} when requiring (i) a new dynamics in (dominantly) left-chiral semi-leptonic transitions, and (ii) a flavour structure implying dominant (but not exclusive) couplings to the third family. The high-$p_T$ phenomenology of the combined solution is very similar to the $R(D^{(*)})$ discussion above. An exceptional working model is $U_1^\mu$ vector LQ. For UV completion as a massive gauge boson see refs.~\cite{DiLuzio:2017vat,Bordone:2017bld}.

\section{Conclusions}
\label{sec:conclusions}

We address the need for an up-to-date Monte Carlo event generator output that can be used for the current and future experimental searches and search recasts concerning scalar and vector LQs. We implement and provide ready-to-use LQ models in the universal \FeynRules output format assuming the SM fermion content and conservation of baryon and lepton numbers for all scalar LQs as well as one vector LQ. Scalar LQ implementations include NLO QCD corrections. We validate our numerical results with the existing NLO calculations for the pair production and present novel results for the single LQ production for scalar (vector) LQs at the NLO (LO) level. The numerical output comprises the NLO QCD inclusive cross sections in proton-proton collisions at 13\,TeV, 14\,TeV, and 27\,TeV center-of-mass energies as a function of the LQ mass. These results can be particularly useful for the current and future LHC data analyses, accurate search recasts, and the flavour dependent studies of the LQ signatures at colliders within the \MadGraph framework. We also discuss aspects of the LQ searches at a hadron collider and outline the high-$p_T$ search strategy for LQs recently proposed in the literature to resolve experimental anomalies in $B$-meson decays.


\acknowledgments
This work has been supported in part by Croatian Science Foundation under the project 7118. I.D.\ would like to thank the CERN Theoretical Physics Department on hospitality, where part of this work was done. We would like to thank Darius A.\ Faroughy for useful discussions.

\appendix
\section{LQ cross sections in proton-proton collisions}
\label{appendix:A}
We present in tables~\ref{tab:13PDF4LHC15},~\ref{tab:14PDF4LHC15}, and~\ref{tab:27PDF4LHC15} total inclusive cross sections in pb for the {\tt PDF4LHC15} PDF sets~\cite{Butterworth:2015oua} as a function of the LQ mass $m_\mathrm{LQ}$ at 13\,TeV, 14\,TeV, and 27\,TeV center-of-mass energies for the proton-proton collisions, respectively. These results are valid in the narrow width approximation. For the discussion on the effects beyond this approximation see ref.~\cite{Hammett:2015sea}.
\begin{table}
{\small
\begin{tabular}{cccc}
\hline
$m_\mathrm{LQ}$\,(TeV)&$\sigma^\mathrm{pair}$ (pb)&$\sigma^\mathrm{single}_u$ (pb)&$\sigma^\mathrm{single}_d$ (pb) \\
\hline
\hline
0.2 & $63.5^{+12.9\%+2.7\%}_{-12.5\%-2.7\%}$ & $116.^{+7.1\%+1.5\%}_{-6.1\%-1.5\%}$ & $76.3^{+7.0\%+2.2\%}_{-6.1\%-2.2\%}$ \\
0.4 & $1.81^{+11.2\%+4.2\%}_{-12.5\%-4.2\%}$ & $7.56^{+7.2\%+1.4\%}_{-6.8\%-1.4\%}$ & $4.49^{+7.3\%+2.2\%}_{-7.0\%-2.2\%}$ \\
0.6 & $0.169^{+11.4\%+5.5\%}_{-13.0\%-5.5\%}$ & $1.31^{+7.3\%+1.7\%}_{-7.3\%-1.7\%}$ & $0.722^{+7.6\%+2.3\%}_{-7.5\%-2.3\%}$ \\
0.8 & $0.0266^{+11.7\%+6.8\%}_{-13.4\%-6.8\%}$ & $0.341^{+7.8\%+2.1\%}_{-7.8\%-2.1\%}$ & $0.177^{+7.9\%+2.5\%}_{-8.0\%-2.5\%}$ \\
1.0 & $0.00553^{+11.3\%+8.3\%}_{-13.4\%-8.3\%}$ & $0.112^{+7.9\%+2.4\%}_{-8.1\%-2.4\%}$ & $0.0552^{+8.2\%+2.8\%}_{-8.3\%-2.8\%}$ \\
1.2 & $0.00134^{+11.8\%+9.9\%}_{-13.9\%-9.9\%}$ & $0.0426^{+8.2\%+2.8\%}_{-8.4\%-2.8\%}$ & $0.02^{+8.7\%+3.1\%}_{-8.8\%-3.1\%}$ \\
1.4 & $0.000367^{+12.1\%+11.8\%}_{-14.2\%-11.8\%}$ & $0.0178^{+8.6\%+3.2\%}_{-8.8\%-3.2\%}$ & $0.00797^{+8.8\%+3.5\%}_{-9.0\%-3.5\%}$ \\
1.6 & $0.000107^{+12.5\%+13.9\%}_{-14.6\%-13.9\%}$ & $0.00808^{+8.7\%+3.6\%}_{-9.0\%-3.6\%}$ & $0.00346^{+9.2\%+3.9\%}_{-9.4\%-3.9\%}$ \\
1.8 & $0.0000329^{+13.4\%+16.4\%}_{-15.2\%-16.4\%}$ & $0.00387^{+9.1\%+4.0\%}_{-9.3\%-4.0\%}$ & $0.00159^{+9.5\%+4.4\%}_{-9.7\%-4.4\%}$ \\
2.0 & $0.0000103^{+14.2\%+19.3\%}_{-15.7\%-19.3\%}$ & $0.00193^{+9.5\%+4.5\%}_{-9.7\%-4.5\%}$ & $0.000761^{+9.8\%+5.0\%}_{-10.0\%-5.0\%}$ \\
2.2 & $(3.29 \times 10^{-6})^{+14.3\%+23.1\%}_{-15.9\%-23.1\%}$ & $0.000993^{+9.6\%+5.0\%}_{-9.8\%-5.0\%}$ & $0.000378^{+10.0\%+5.6\%}_{-10.2\%-5.6\%}$ \\
2.4 & $(1.08 \times 10^{-6})^{+15.0\%+28.8\%}_{-16.3\%-28.8\%}$ & $0.000532^{+9.7\%+5.6\%}_{-10.0\%-5.6\%}$ & $0.000195^{+10.2\%+6.3\%}_{-10.4\%-6.3\%}$ \\
2.6 & $(3.47 \times 10^{-7})^{+15.5\%+37.7\%}_{-16.6\%-37.7\%}$ & $0.000289^{+10.2\%+6.2\%}_{-10.4\%-6.2\%}$ & $0.000102^{+10.7\%+7.1\%}_{-10.8\%-7.1\%}$ \\
2.8 & $(1.12 \times 10^{-7})^{+16.0\%+54.1\%}_{-16.8\%-54.1\%}$ & $0.000159^{+10.5\%+6.9\%}_{-10.7\%-6.9\%}$ & $0.0000542^{+11.1\%+7.9\%}_{-11.1\%-7.9\%}$ \\
3.0 & $(3.69 \times 10^{-8})^{+15.8\%+83.8\%}_{-16.6\%-83.8\%}$ & $0.0000894^{+11.0\%+7.5\%}_{-11.0\%-7.5\%}$ & $0.0000296^{+11.3\%+8.9\%}_{-11.3\%-8.9\%}$ \\
\hline
\hline
$m_\mathrm{LQ}$\,(TeV)&$\sigma^\mathrm{single}_s$ (pb)&$\sigma^\mathrm{single}_c$ (pb)&$\sigma^\mathrm{single}_b$ (pb) \\
\hline
\hline
0.2 & $28.5^{+7.4\%+8.4\%}_{-6.5\%-8.4\%}$ & $19.8^{+8.0\%+3.3\%}_{-7.4\%-3.3\%}$ & $12.3^{+8.6\%+4.3\%}_{-9.3\%-4.3\%}$ \\
0.4 & $1.31^{+6.6\%+10.0\%}_{-6.4\%-10.0\%}$ & $0.846^{+6.0\%+4.6\%}_{-5.2\%-4.6\%}$ & $0.537^{+7.4\%+5.4\%}_{-6.6\%-5.4\%}$ \\
0.6 & $0.175^{+6.9\%+11.7\%}_{-7.0\%-11.7\%}$ & $0.107^{+6.1\%+5.9\%}_{-6.1\%-5.9\%}$ & $0.0672^{+5.7\%+6.5\%}_{-4.7\%-6.5\%}$ \\
0.8 & $0.0368^{+7.4\%+13.7\%}_{-7.6\%-13.7\%}$ & $0.0215^{+6.6\%+7.4\%}_{-6.7\%-7.4\%}$ & $0.0134^{+5.2\%+7.6\%}_{-4.8\%-7.6\%}$ \\
1.0 & $0.0101^{+7.7\%+16.1\%}_{-8.0\%-16.1\%}$ & $0.00568^{+6.6\%+9.0\%}_{-6.9\%-9.0\%}$ & $0.00351^{+4.8\%+9.0\%}_{-5.3\%-9.0\%}$ \\
1.2 & $0.00328^{+8.1\%+19.2\%}_{-8.3\%-19.2\%}$ & $0.00177^{+6.9\%+10.6\%}_{-7.3\%-10.6\%}$ & $0.00109^{+5.1\%+10.3\%}_{-5.7\%-10.3\%}$ \\
1.4 & $0.0012^{+8.4\%+22.7\%}_{-8.7\%-22.7\%}$ & $0.000617^{+7.3\%+12.4\%}_{-7.8\%-12.4\%}$ & $0.000379^{+5.6\%+11.9\%}_{-6.2\%-11.9\%}$ \\
1.6 & $0.000476^{+8.6\%+27.0\%}_{-9.0\%-27.0\%}$ & $0.000237^{+7.7\%+14.3\%}_{-8.1\%-14.3\%}$ & $0.000145^{+5.7\%+13.5\%}_{-6.4\%-13.5\%}$ \\
1.8 & $0.000202^{+8.9\%+31.4\%}_{-9.3\%-31.4\%}$ & $0.000097^{+7.9\%+16.3\%}_{-8.3\%-16.3\%}$ & $0.0000587^{+5.9\%+15.3\%}_{-6.7\%-15.3\%}$ \\
2.0 & $0.0000916^{+9.2\%+37.1\%}_{-9.5\%-37.1\%}$ & $0.0000413^{+8.2\%+18.6\%}_{-8.7\%-18.6\%}$ & $0.000025^{+6.3\%+17.3\%}_{-7.1\%-17.3\%}$ \\
2.2 & $0.000043^{+9.6\%+43.2\%}_{-9.9\%-43.2\%}$ & $0.0000187^{+8.5\%+20.8\%}_{-9.0\%-20.8\%}$ & $0.0000112^{+6.8\%+19.3\%}_{-7.5\%-19.3\%}$ \\
2.4 & $0.000021^{+9.9\%+49.6\%}_{-10.1\%-49.6\%}$ & $(8.68 \times 10^{-6})^{+9.1\%+23.3\%}_{-9.5\%-23.3\%}$ & $(5.16 \times 10^{-6})^{+7.0\%+21.6\%}_{-7.8\%-21.6\%}$ \\
2.6 & $0.0000105^{+10.0\%+57.4\%}_{-10.3\%-57.4\%}$ & $(4.1 \times 10^{-6})^{+9.3\%+26.0\%}_{-9.7\%-26.0\%}$ & $(2.45 \times 10^{-6})^{+7.1\%+24.0\%}_{-8.0\%-24.0\%}$ \\
2.8 & $(5.49 \times 10^{-6})^{+10.4\%+64.3\%}_{-10.6\%-64.3\%}$ & $(1.99 \times 10^{-6})^{+9.6\%+28.8\%}_{-10.0\%-28.8\%}$ & $(1.18 \times 10^{-6})^{+7.6\%+26.5\%}_{-8.4\%-26.5\%}$ \\
3.0 & $(2.91 \times 10^{-6})^{+10.7\%+72.8\%}_{-10.8\%-72.8\%}$ & $(9.96 \times 10^{-7})^{+9.9\%+31.8\%}_{-10.2\%-31.8\%}$ & $(5.81 \times 10^{-7})^{+7.8\%+29.2\%}_{-8.6\%-29.2\%}$ \\
\hline
\end{tabular}}
\caption{\label{tab:13PDF4LHC15} Total inclusive cross sections in pb for the {\tt PDF4LHC15} PDF sets~\cite{Butterworth:2015oua} as a function of the LQ mass at 13\,TeV center-of-mass energy for the proton-proton collisions. $\sigma^\mathrm{pair}$ corresponds to the LQ pair production. $\sigma^\mathrm{single}_{u,d,s,c,b}$ describe single LQ production cross sections through corresponding quark flavour when the associated Yukawa coupling strength is set to one. The cross section dependancy on the change in the renormalisation ($\mu_R$) and factorisation ($\mu_F$) scales is taken into account through the following scale variations: $\mu_R=\mu_F=m_\mathrm{LQ}/2, m_\mathrm{LQ}, 2 m_\mathrm{LQ}$. First (second) uncertainty is due to the renormalisation $\mu_R$ and factorisation $\mu_F$ scale (PDF) variations and is given in per cent units.}
\end{table}

\begin{table}[tbp]
{\small
\centering
\begin{tabular}{cccc}
\hline
$m_\mathrm{LQ}$\,(TeV)&$\sigma^\mathrm{pair}$ (pb)&$\sigma^\mathrm{single}_u$ (pb)&$\sigma^\mathrm{single}_d$ (pb) \\
\hline
\hline
0.2 & $75.1^{+12.3\%+2.6\%}_{-12.1\%-2.6\%}$ & $130.^{+7.1\%+1.6\%}_{-6.2\%-1.6\%}$ & $86.8^{+7.3\%+2.2\%}_{-6.2\%-2.2\%}$ \\
0.4 & $2.25^{+11.5\%+4.0\%}_{-12.5\%-4.0\%}$ & $8.68^{+7.6\%+1.4\%}_{-7.0\%-1.4\%}$ & $5.23^{+7.3\%+2.2\%}_{-6.9\%-2.2\%}$ \\
0.6 & $0.222^{+11.1\%+5.2\%}_{-12.7\%-5.2\%}$ & $1.54^{+7.3\%+1.6\%}_{-7.2\%-1.6\%}$ & $0.864^{+7.4\%+2.3\%}_{-7.3\%-2.3\%}$ \\
0.8 & $0.037^{+10.9\%+6.4\%}_{-12.9\%-6.4\%}$ & $0.414^{+7.7\%+2.0\%}_{-7.6\%-2.0\%}$ & $0.218^{+7.8\%+2.5\%}_{-7.8\%-2.5\%}$ \\
1.0 & $0.00787^{+11.5\%+7.7\%}_{-13.4\%-7.7\%}$ & $0.138^{+7.5\%+2.3\%}_{-7.7\%-2.3\%}$ & $0.0686^{+8.0\%+2.7\%}_{-8.1\%-2.7\%}$ \\
1.2 & $0.00204^{+11.5\%+9.1\%}_{-13.6\%-9.1\%}$ & $0.0535^{+8.0\%+2.6\%}_{-8.2\%-2.6\%}$ & $0.0256^{+8.2\%+2.9\%}_{-8.5\%-2.9\%}$ \\
1.4 & $0.000574^{+12.1\%+10.8\%}_{-14.0\%-10.8\%}$ & $0.023^{+8.2\%+3.0\%}_{-8.5\%-3.0\%}$ & $0.0105^{+8.6\%+3.3\%}_{-8.8\%-3.3\%}$ \\
1.6 & $0.000177^{+12.1\%+12.5\%}_{-14.2\%-12.5\%}$ & $0.0107^{+8.4\%+3.3\%}_{-8.7\%-3.3\%}$ & $0.00466^{+8.8\%+3.6\%}_{-9.1\%-3.6\%}$ \\
1.8 & $0.0000585^{+13.2\%+14.7\%}_{-14.9\%-14.7\%}$ & $0.00518^{+8.6\%+3.7\%}_{-9.0\%-3.7\%}$ & $0.00219^{+9.3\%+4.1\%}_{-9.5\%-4.1\%}$ \\
2.0 & $0.0000193^{+13.3\%+17.0\%}_{-15.1\%-17.0\%}$ & $0.00267^{+9.1\%+4.2\%}_{-9.4\%-4.2\%}$ & $0.00108^{+9.5\%+4.6\%}_{-9.7\%-4.6\%}$ \\
2.2 & $(6.72 \times 10^{-6})^{+14.1\%+19.9\%}_{-15.6\%-19.9\%}$ & $0.0014^{+9.3\%+4.6\%}_{-9.6\%-4.6\%}$ & $0.000548^{+9.7\%+5.1\%}_{-9.9\%-5.1\%}$ \\
2.4 & $(2.33 \times 10^{-6})^{+14.5\%+23.8\%}_{-15.9\%-23.8\%}$ & $0.00077^{+9.6\%+5.1\%}_{-9.8\%-5.1\%}$ & $0.000288^{+10.1\%+5.7\%}_{-10.3\%-5.7\%}$ \\
2.6 & $(8.2 \times 10^{-7})^{+15.1\%+29.0\%}_{-16.3\%-29.0\%}$ & $0.000427^{+9.5\%+5.7\%}_{-9.9\%-5.7\%}$ & $0.000156^{+10.3\%+6.4\%}_{-10.4\%-6.4\%}$ \\
2.8 & $(2.87 \times 10^{-7})^{+15.3\%+37.7\%}_{-16.5\%-37.7\%}$ & $0.000241^{+10.0\%+6.2\%}_{-10.3\%-6.2\%}$ & $0.0000851^{+10.6\%+7.1\%}_{-10.8\%-7.1\%}$ \\
3.0 & $(1.02 \times 10^{-7})^{+15.8\%+52.5\%}_{-16.6\%-52.5\%}$ & $0.000139^{+10.4\%+6.8\%}_{-10.5\%-6.8\%}$ & $0.0000477^{+10.9\%+7.9\%}_{-11.0\%-7.9\%}$ \\
\hline
$m_\mathrm{LQ}$\,(TeV)&$\sigma^\mathrm{single}_s$ (pb)&$\sigma^\mathrm{single}_c$ (pb)&$\sigma^\mathrm{single}_b$ (pb) \\
\hline
\hline
0.2 & $33.1^{+7.4\%+8.3\%}_{-6.6\%-8.3\%}$ & $23.2^{+8.3\%+3.1\%}_{-7.8\%-3.1\%}$ & $14.7^{+9.4\%+4.2\%}_{-9.9\%-4.2\%}$ \\
0.4 & $1.57^{+6.3\%+9.7\%}_{-6.1\%-9.7\%}$ & $1.04^{+6.2\%+4.4\%}_{-5.2\%-4.4\%}$ & $0.653^{+7.2\%+5.2\%}_{-6.6\%-5.2\%}$ \\
0.6 & $0.216^{+6.8\%+11.2\%}_{-6.8\%-11.2\%}$ & $0.135^{+5.9\%+5.6\%}_{-5.9\%-5.6\%}$ & $0.085^{+6.2\%+6.2\%}_{-5.1\%-6.2\%}$ \\
0.8 & $0.0471^{+7.1\%+13.0\%}_{-7.3\%-13.0\%}$ & $0.028^{+6.2\%+7.0\%}_{-6.4\%-7.0\%}$ & $0.0175^{+5.2\%+7.2\%}_{-4.7\%-7.2\%}$ \\
1.0 & $0.0134^{+7.5\%+15.1\%}_{-7.7\%-15.1\%}$ & $0.00757^{+6.7\%+8.4\%}_{-6.9\%-8.4\%}$ & $0.00469^{+4.9\%+8.5\%}_{-5.1\%-8.5\%}$ \\
1.2 & $0.00442^{+7.8\%+17.7\%}_{-8.1\%-17.7\%}$ & $0.00244^{+6.9\%+9.9\%}_{-7.2\%-9.9\%}$ & $0.00149^{+5.0\%+9.7\%}_{-5.5\%-9.7\%}$ \\
1.4 & $0.00165^{+8.1\%+20.6\%}_{-8.4\%-20.6\%}$ & $0.000868^{+7.2\%+11.6\%}_{-7.5\%-11.6\%}$ & $0.000535^{+5.3\%+11.0\%}_{-5.9\%-11.0\%}$ \\
1.6 & $0.000676^{+8.4\%+24.4\%}_{-8.7\%-24.4\%}$ & $0.000344^{+7.5\%+13.2\%}_{-7.9\%-13.2\%}$ & $0.000211^{+5.5\%+12.6\%}_{-6.2\%-12.6\%}$ \\
1.8 & $0.000294^{+8.8\%+28.5\%}_{-9.1\%-28.5\%}$ & $0.000144^{+7.7\%+15.0\%}_{-8.2\%-15.0\%}$ & $0.0000884^{+5.8\%+14.2\%}_{-6.5\%-14.2\%}$ \\
2.0 & $0.000135^{+9.1\%+33.2\%}_{-9.4\%-33.2\%}$ & $0.000064^{+7.7\%+16.8\%}_{-8.3\%-16.8\%}$ & $0.0000388^{+6.0\%+15.8\%}_{-6.8\%-15.8\%}$ \\
2.2 & $0.0000653^{+9.3\%+38.5\%}_{-9.6\%-38.5\%}$ & $0.0000293^{+8.3\%+18.9\%}_{-8.7\%-18.9\%}$ & $0.0000179^{+6.4\%+17.6\%}_{-7.1\%-17.6\%}$ \\
2.4 & $0.0000327^{+9.4\%+44.1\%}_{-9.7\%-44.1\%}$ & $0.000014^{+8.4\%+21.2\%}_{-8.9\%-21.2\%}$ & $(8.38 \times 10^{-6})^{+6.7\%+19.7\%}_{-7.5\%-19.7\%}$ \\
2.6 & $0.0000167^{+9.7\%+50.6\%}_{-10.0\%-50.6\%}$ & $(6.93 \times 10^{-6})^{+8.7\%+23.4\%}_{-9.2\%-23.4\%}$ & $(4.13 \times 10^{-6})^{+7.1\%+21.8\%}_{-7.8\%-21.8\%}$ \\
2.8 & $(8.82 \times 10^{-6})^{+10.0\%+57.0\%}_{-10.2\%-57.0\%}$ & $(3.47 \times 10^{-6})^{+9.1\%+25.9\%}_{-9.6\%-25.9\%}$ & $(2.06 \times 10^{-6})^{+7.1\%+23.8\%}_{-7.9\%-23.8\%}$ \\
3.0 & $(4.79 \times 10^{-6})^{+10.3\%+63.8\%}_{-10.5\%-63.8\%}$ & $(1.78 \times 10^{-6})^{+9.3\%+28.4\%}_{-9.8\%-28.4\%}$ & $(1.05 \times 10^{-6})^{+7.5\%+26.4\%}_{-8.3\%-26.4\%}$ \\
\hline
\end{tabular} }
\caption{\label{tab:14PDF4LHC15} Total inclusive cross sections in pb for the {\tt PDF4LHC15} PDF sets~\cite{Butterworth:2015oua} as a function of the LQ mass at 14\,TeV center-of-mass energy for the proton-proton collisions. $\sigma^\mathrm{pair}$ corresponds to the LQ pair production. $\sigma^\mathrm{single}_{u,d,s,c,b}$ describe single LQ productions through corresponding quark flavour when the associated Yukawa coupling strength is set to one. The cross section dependancy on the change in the renormalisation ($\mu_R$) and factorisation ($\mu_F$) scales is taken into account through the following scale variations: $\mu_R=\mu_F=m_\mathrm{LQ}/2, m_\mathrm{LQ}, 2 m_\mathrm{LQ}$. First (second) uncertainty is due to the renormalisation $\mu_R$ and factorisation $\mu_F$ scale (PDF) variations and is given in per cent units.}
\end{table}

\begin{table}[tbp]
{\small
\centering
\begin{tabular}{cccc}
\hline
$m_\mathrm{LQ}$\,(TeV)&$\sigma^\mathrm{pair}$ (pb)&$\sigma^\mathrm{single}_u$ (pb)&$\sigma^\mathrm{single}_d$ (pb) \\
\hline
\hline
1.0 & $0.106^{+9.6\%+4.7\%}_{-11.3\%-4.7\%}$ & $0.676^{+6.6\%+1.5\%}_{-6.5\%-1.5\%}$ & $0.388^{+6.6\%+2.2\%}_{-6.6\%-2.2\%}$ \\
1.4 & $0.0138^{+9.8\%+5.9\%}_{-11.7\%-5.9\%}$ & $0.149^{+6.6\%+1.9\%}_{-6.7\%-1.9\%}$ & $0.0797^{+6.9\%+2.4\%}_{-7.0\%-2.4\%}$ \\
1.8 & $0.00257^{+9.8\%+7.1\%}_{-11.9\%-7.1\%}$ & $0.0442^{+6.9\%+2.2\%}_{-7.1\%-2.2\%}$ & $0.0224^{+7.2\%+2.6\%}_{-7.4\%-2.6\%}$ \\
2.2 & $0.00059^{+10.4\%+8.5\%}_{-12.4\%-8.5\%}$ & $0.0158^{+7.3\%+2.6\%}_{-7.5\%-2.6\%}$ & $0.0076^{+7.4\%+2.9\%}_{-7.7\%-2.9\%}$ \\
2.6 & $0.000159^{+10.5\%+10.0\%}_{-12.6\%-10.0\%}$ & $0.00635^{+7.4\%+2.9\%}_{-7.7\%-2.9\%}$ & $0.00289^{+7.7\%+3.2\%}_{-8.0\%-3.2\%}$ \\
3.0 & $0.0000453^{+10.8\%+11.6\%}_{-12.9\%-11.6\%}$ & $0.00279^{+7.6\%+3.3\%}_{-7.9\%-3.3\%}$ & $0.00122^{+7.9\%+3.6\%}_{-8.3\%-3.6\%}$ \\
3.4 & $0.0000139^{+11.0\%+13.6\%}_{-13.2\%-13.6\%}$ & $0.00131^{+7.8\%+3.7\%}_{-8.2\%-3.7\%}$ & $0.000548^{+8.1\%+4.0\%}_{-8.5\%-4.0\%}$ \\
3.8 & $(4.44 \times 10^{-6})^{+11.8\%+16.0\%}_{-13.7\%-16.0\%}$ & $0.000643^{+8.0\%+4.1\%}_{-8.4\%-4.1\%}$ & $0.000261^{+8.4\%+4.5\%}_{-8.8\%-4.5\%}$ \\
4.2 & $(1.45 \times 10^{-6})^{+12.2\%+18.8\%}_{-14.0\%-18.8\%}$ & $0.000332^{+8.3\%+4.6\%}_{-8.7\%-4.6\%}$ & $0.000129^{+8.7\%+5.1\%}_{-9.1\%-5.1\%}$ \\
4.6 & $(4.82 \times 10^{-7})^{+12.6\%+22.8\%}_{-14.4\%-22.8\%}$ & $0.000175^{+8.6\%+5.1\%}_{-8.9\%-5.1\%}$ & $0.0000657^{+8.8\%+5.6\%}_{-9.2\%-5.6\%}$ \\
5.0 & $(1.62 \times 10^{-7})^{+13.1\%+28.7\%}_{-14.7\%-28.7\%}$ & $0.0000947^{+8.6\%+5.6\%}_{-9.1\%-5.6\%}$ & $0.0000343^{+9.2\%+6.3\%}_{-9.5\%-6.3\%}$ \\
\hline
$m_\mathrm{LQ}$ (TeV)&$\sigma^\mathrm{single}_s$ (pb)&$\sigma^\mathrm{single}_c$ (pb)&$\sigma^\mathrm{single}_b$ (pb) \\
\hline
\hline
1.0 & $0.105^{+6.0\%+10.4\%}_{-6.0\%-10.4\%}$ & $0.0683^{+5.2\%+5.1\%}_{-5.3\%-5.1\%}$ & $0.0445^{+5.6\%+5.7\%}_{-4.7\%-5.7\%}$ \\
1.4 & $0.0183^{+6.4\%+12.3\%}_{-6.6\%-12.3\%}$ & $0.0113^{+5.5\%+6.4\%}_{-5.7\%-6.4\%}$ & $0.0073^{+4.9\%+6.7\%}_{-4.2\%-6.7\%}$ \\
1.8 & $0.00449^{+6.6\%+14.4\%}_{-6.9\%-14.4\%}$ & $0.00263^{+5.9\%+7.9\%}_{-6.2\%-7.9\%}$ & $0.00169^{+4.3\%+7.9\%}_{-4.7\%-7.9\%}$ \\
2.2 & $0.00136^{+7.0\%+17.0\%}_{-7.3\%-17.0\%}$ & $0.000771^{+6.1\%+9.5\%}_{-6.5\%-9.5\%}$ & $0.000487^{+4.5\%+9.2\%}_{-5.0\%-9.2\%}$ \\
2.6 & $0.000472^{+7.3\%+20.1\%}_{-7.7\%-20.1\%}$ & $0.000256^{+6.3\%+11.1\%}_{-6.8\%-11.1\%}$ & $0.000162^{+5.1\%+10.6\%}_{-5.6\%-10.6\%}$ \\
3.0 & $0.000182^{+7.4\%+23.7\%}_{-7.8\%-23.7\%}$ & $0.0000948^{+6.7\%+12.8\%}_{-7.2\%-12.8\%}$ & $0.0000595^{+5.2\%+12.1\%}_{-5.8\%-12.1\%}$ \\
3.4 & $0.0000755^{+7.6\%+27.8\%}_{-8.1\%-27.8\%}$ & $0.000038^{+6.7\%+14.6\%}_{-7.3\%-14.6\%}$ & $0.0000237^{+5.3\%+13.7\%}_{-6.0\%-13.7\%}$ \\
3.8 & $0.0000335^{+8.1\%+32.8\%}_{-8.5\%-32.8\%}$ & $0.0000161^{+7.1\%+16.5\%}_{-7.7\%-16.5\%}$ & $(0.00001 \times 10^{})^{+5.9\%+15.4\%}_{-6.5\%-15.4\%}$ \\
4.2 & $0.0000155^{+8.3\%+38.0\%}_{-8.7\%-38.0\%}$ & $(7.17 \times 10^{-6})^{+7.6\%+18.6\%}_{-8.1\%-18.6\%}$ & $(4.38 \times 10^{-6})^{+6.1\%+17.2\%}_{-6.8\%-17.2\%}$ \\
4.6 & $(7.51 \times 10^{-6})^{+8.5\%+44.1\%}_{-8.9\%-44.1\%}$ & $(3.28 \times 10^{-6})^{+7.8\%+20.8\%}_{-8.3\%-20.8\%}$ & $(2.03 \times 10^{-6})^{+6.2\%+19.2\%}_{-6.9\%-19.2\%}$ \\
5.0 & $(3.74 \times 10^{-6})^{+8.9\%+50.6\%}_{-9.2\%-50.6\%}$ & $(1.56 \times 10^{-6})^{+8.0\%+23.1\%}_{-8.5\%-23.1\%}$ & $(9.55 \times 10^{-7})^{+6.5\%+21.3\%}_{-7.2\%-21.3\%}$ \\
\hline
\end{tabular} }
\caption{\label{tab:27PDF4LHC15} Total inclusive cross sections in pb for the {\tt PDF4LHC15} PDF sets~\cite{Butterworth:2015oua} as a function of the LQ mass at 27\,TeV center-of-mass energy for the proton-proton collisions. $\sigma^\mathrm{pair}$ corresponds to the LQ pair production. $\sigma^\mathrm{single}_{u,d,s,c,b}$ describe single LQ productions through corresponding quark flavour when the associated Yukawa coupling strength is set to one. The cross section dependancy on the change in the renormalisation ($\mu_R$) and factorisation ($\mu_F$) scales is taken into account through the following scale variations: $\mu_R=\mu_F=m_\mathrm{LQ}/2, m_\mathrm{LQ}, 2 m_\mathrm{LQ}$. First (second) uncertainty is due to the renormalisation $\mu_R$ and factorisation $\mu_F$ scale (PDF) variations and is given in per cent units.}
\end{table}



\begin{thebibliography}{00}  

\bibitem{Pati:1973uk} 
  J.~C.~Pati and A.~Salam,
  Phys.\ Rev.\ D {\bf 8}, 1240 (1973).
  doi:10.1103/PhysRevD.8.1240


\bibitem{Davidson:1993qk} 
  S.~Davidson, D.~C.~Bailey and B.~A.~Campbell,
  Z.\ Phys.\ C {\bf 61}, 613 (1994)
  doi:10.1007/BF01552629
  [hep-ph/9309310].


\bibitem{Hewett:1997ce} 
  J.~L.~Hewett and T.~G.~Rizzo,
  Phys.\ Rev.\ D {\bf 56}, 5709 (1997)
  doi:10.1103/PhysRevD.56.5709
  [hep-ph/9703337].


\bibitem{Nath:2006ut} 
  P.~Nath and P.~Fileviez Perez,
  Phys.\ Rept.\  {\bf 441}, 191 (2007)
  doi:10.1016/j.physrep.2007.02.010
  [hep-ph/0601023].


\bibitem{Dorsner:2016wpm} 
  I.~Dor\v sner, S.~Fajfer, A.~Greljo, J.~F.~Kamenik and N.~Ko\v snik,
  Phys.\ Rept.\  {\bf 641}, 1 (2016)
  doi:10.1016/j.physrep.2016.06.001
  [arXiv:1603.04993 [hep-ph]].


\bibitem{Dorsner:2014axa} 
  I.~Dorsner, S.~Fajfer and A.~Greljo,
  JHEP {\bf 1410}, 154 (2014)
  doi:10.1007/JHEP10(2014)154
  [arXiv:1406.4831 [hep-ph]].


\bibitem{Mandal:2015vfa} 
  T.~Mandal, S.~Mitra and S.~Seth,
  JHEP {\bf 1507}, 028 (2015)
  doi:10.1007/JHEP07(2015)028
  [arXiv:1503.04689 [hep-ph]].


\bibitem{Diaz:2017lit} 
  B.~Diaz, M.~Schmaltz and Y.~M.~Zhong,
  JHEP {\bf 1710}, 097 (2017)
  doi:10.1007/JHEP10(2017)097
  [arXiv:1706.05033 [hep-ph]].


\bibitem{Bandyopadhyay:2018syt} 
  P.~Bandyopadhyay and R.~Mandal,
  arXiv:1801.04253 [hep-ph].


\bibitem{Kramer:2004df} 
  M.~Kramer, T.~Plehn, M.~Spira and P.~M.~Zerwas,
  Phys.\ Rev.\ D {\bf 71}, 057503 (2005)
  doi:10.1103/PhysRevD.71.057503
  [hep-ph/0411038].


\bibitem{Mandal:2015lca} 
  T.~Mandal, S.~Mitra and S.~Seth,
  Phys.\ Rev.\ D {\bf 93}, no. 3, 035018 (2016)
  doi:10.1103/PhysRevD.93.035018
  [arXiv:1506.07369 [hep-ph]].

\bibitem{Alves:2002tj} 
  A.~Alves, O.~Eboli and T.~Plehn,
  Phys.\ Lett.\ B {\bf 558}, 165 (2003)
  doi:10.1016/S0370-2693(03)00266-1
  [hep-ph/0211441].
  
\bibitem{Hammett:2015sea} 
  J.~B.~Hammett and D.~A.~Ross,
  JHEP {\bf 1507}, 148 (2015)
  doi:10.1007/JHEP07(2015)148
  [arXiv:1501.06719 [hep-ph]].


\bibitem{Buttazzo:2017ixm} 
  D.~Buttazzo, A.~Greljo, G.~Isidori and D.~Marzocca,
  JHEP {\bf 1711}, 044 (2017)
  doi:10.1007/JHEP11(2017)044
  [arXiv:1706.07808 [hep-ph]].


\bibitem{Buchmuller:1986zs} 
  W.~Buchmuller, R.~Ruckl and D.~Wyler,
  Phys.\ Lett.\ B {\bf 191}, 442 (1987)
  Erratum: [Phys.\ Lett.\ B {\bf 448}, 320 (1999)].
  doi:10.1016/S0370-2693(99)00014-3, 10.1016/0370-2693(87)90637-X


\bibitem{Alloul:2013bka} 
  A.~Alloul, N.~D.~Christensen, C.~Degrande, C.~Duhr and B.~Fuks,
  Comput.\ Phys.\ Commun.\  {\bf 185}, 2250 (2014)
  doi:10.1016/j.cpc.2014.04.012
  [arXiv:1310.1921 [hep-ph]].


\bibitem{Alwall:2014hca} 
  J.~Alwall {\it et al.},
  JHEP {\bf 1407}, 079 (2014)
  doi:10.1007/JHEP07(2014)079
  [arXiv:1405.0301 [hep-ph]].


\bibitem{Degrande:2014vpa} 
  C.~Degrande,
  Comput.\ Phys.\ Commun.\  {\bf 197}, 239 (2015)
  doi:10.1016/j.cpc.2015.08.015
  [arXiv:1406.3030 [hep-ph]].


\bibitem{Hahn:2000kx} 
  T.~Hahn,
  Comput.\ Phys.\ Commun.\  {\bf 140}, 418 (2001)
  doi:10.1016/S0010-4655(01)00290-9
  [hep-ph/0012260].


\bibitem{Hirschi:2011pa} 
  V.~Hirschi, R.~Frederix, S.~Frixione, M.~V.~Garzelli, F.~Maltoni and R.~Pittau,
  JHEP {\bf 1105}, 044 (2011)
  doi:10.1007/JHEP05(2011)044
  [arXiv:1103.0621 [hep-ph]].


\bibitem{Peraro:2014cba} 
  T.~Peraro,
  Comput.\ Phys.\ Commun.\  {\bf 185}, 2771 (2014)
  doi:10.1016/j.cpc.2014.06.017
  [arXiv:1403.1229 [hep-ph]].


\bibitem{Hirschi:2016mdz} 
  V.~Hirschi and T.~Peraro,
  JHEP {\bf 1606}, 060 (2016)
  doi:10.1007/JHEP06(2016)060
  [arXiv:1604.01363 [hep-ph]].


\bibitem{Frixione:1995ms} 
  S.~Frixione, Z.~Kunszt and A.~Signer,
  Nucl.\ Phys.\ B {\bf 467}, 399 (1996)
  doi:10.1016/0550-3213(96)00110-1
  [hep-ph/9512328].


\bibitem{Frederix:2009yq} 
  R.~Frederix, S.~Frixione, F.~Maltoni and T.~Stelzer,
  JHEP {\bf 0910}, 003 (2009)
  doi:10.1088/1126-6708/2009/10/003
  [arXiv:0908.4272 [hep-ph]].


\bibitem{Plehn:1997az} 
  T.~Plehn, H.~Spiesberger, M.~Spira and P.~M.~Zerwas,
  Z.\ Phys.\ C {\bf 74}, 611 (1997)
  doi:10.1007/s002880050426
  [hep-ph/9703433].


\bibitem{DiLuzio:2017vat} 
  L.~Di Luzio, A.~Greljo and M.~Nardecchia,
  Phys.\ Rev.\ D {\bf 96}, no. 11, 115011 (2017)
  doi:10.1103/PhysRevD.96.115011
  [arXiv:1708.08450 [hep-ph]].


\bibitem{Bordone:2017bld} 
  M.~Bordone, C.~Cornella, J.~Fuentes-Martin and G.~Isidori,
  arXiv:1712.01368 [hep-ph].


\bibitem{Barbieri:2017tuq} 
  R.~Barbieri and A.~Tesi,
  arXiv:1712.06844 [hep-ph].


\bibitem{Butterworth:2015oua} 
  J.~Butterworth {\it et al.},
  J.\ Phys.\ G {\bf 43}, 023001 (2016)
  doi:10.1088/0954-3899/43/2/023001
  [arXiv:1510.03865 [hep-ph]].


\bibitem{Guena:2004sq} 
  J.~Guena, M.~Lintz and M.~A.~Bouchiat,
  Phys.\ Rev.\ A {\bf 71}, 042108 (2005)
  doi:10.1103/PhysRevA.71.042108
  [physics/0412017 [physics.atom-ph]].


\bibitem{Wood:1997zq} 
  C.~S.~Wood, S.~C.~Bennett, D.~Cho, B.~P.~Masterson, J.~L.~Roberts, C.~E.~Tanner and C.~E.~Wieman,
  Science {\bf 275}, 1759 (1997).
  doi:10.1126/science.275.5307.1759


\bibitem{Blumlein:1996qp} 
  J.~Blumlein, E.~Boos and A.~Kryukov,
  Z.\ Phys.\ C {\bf 76}, 137 (1997)
  doi:10.1007/s002880050538
  [hep-ph/9610408].


\bibitem{Ball:2014uwa} 
  R.~D.~Ball {\it et al.} [NNPDF Collaboration],
  JHEP {\bf 1504}, 040 (2015)
  doi:10.1007/JHEP04(2015)040
  [arXiv:1410.8849 [hep-ph]].


\bibitem{Pumplin:2002vw} 
  J.~Pumplin, D.~R.~Stump, J.~Huston, H.~L.~Lai, P.~M.~Nadolsky and W.~K.~Tung,
  JHEP {\bf 0207}, 012 (2002)
  doi:10.1088/1126-6708/2002/07/012
  [hep-ph/0201195].


\bibitem{Lees:2013uzd} 
  J.~P.~Lees {\it et al.} [BaBar Collaboration],
  Phys.\ Rev.\ D {\bf 88}, no. 7, 072012 (2013)
  doi:10.1103/PhysRevD.88.072012
  [arXiv:1303.0571 [hep-ex]].


\bibitem{Hirose:2016wfn} 
  S.~Hirose {\it et al.} [Belle Collaboration],
  Phys.\ Rev.\ Lett.\  {\bf 118}, no. 21, 211801 (2017)
  doi:10.1103/PhysRevLett.118.211801
  [arXiv:1612.00529 [hep-ex]].


\bibitem{Aaij:2015yra} 
  R.~Aaij {\it et al.} [LHCb Collaboration],
  Phys.\ Rev.\ Lett.\  {\bf 115}, no. 11, 111803 (2015)
  Erratum: [Phys.\ Rev.\ Lett.\  {\bf 115}, no. 15, 159901 (2015)]
  doi:10.1103/PhysRevLett.115.159901, 10.1103/PhysRevLett.115.111803
  [arXiv:1506.08614 [hep-ex]].


\bibitem{Aaij:2014ora} 
  R.~Aaij {\it et al.} [LHCb Collaboration],
  Phys.\ Rev.\ Lett.\  {\bf 113}, 151601 (2014)
  doi:10.1103/PhysRevLett.113.151601
  [arXiv:1406.6482 [hep-ex]].


\bibitem{Aaij:2017vbb} 
  R.~Aaij {\it et al.} [LHCb Collaboration],
  JHEP {\bf 1708}, 055 (2017)
  doi:10.1007/JHEP08(2017)055
  [arXiv:1705.05802 [hep-ex]].


\bibitem{Aaij:2013qta} 
  R.~Aaij {\it et al.} [LHCb Collaboration],
  Phys.\ Rev.\ Lett.\  {\bf 111}, 191801 (2013)
  doi:10.1103/PhysRevLett.111.191801
  [arXiv:1308.1707 [hep-ex]].


\bibitem{Aaij:2015oid} 
  R.~Aaij {\it et al.} [LHCb Collaboration],
  JHEP {\bf 1602}, 104 (2016)
  doi:10.1007/JHEP02(2016)104
  [arXiv:1512.04442 [hep-ex]].


\bibitem{Capdevila:2017bsm} 
  B.~Capdevila, A.~Crivellin, S.~Descotes-Genon, J.~Matias and J.~Virto,
  arXiv:1704.05340 [hep-ph].


\bibitem{Altmannshofer:2017yso} 
  W.~Altmannshofer, P.~Stangl and D.~M.~Straub,
  Phys.\ Rev.\ D {\bf 96}, no. 5, 055008 (2017)
  doi:10.1103/PhysRevD.96.055008
  [arXiv:1704.05435 [hep-ph]].


\bibitem{Geng:2017svp} 
  L.~S.~Geng, B.~Grinstein, S.~Jager, J.~Martin Camalich, X.~L.~Ren and R.~X.~Shi,
  Phys.\ Rev.\ D {\bf 96}, no. 9, 093006 (2017)
  doi:10.1103/PhysRevD.96.093006
  [arXiv:1704.05446 [hep-ph]].

\bibitem{DAmico:2017mtc} 
  G.~D'Amico, M.~Nardecchia, P.~Panci, F.~Sannino, A.~Strumia, R.~Torre and A.~Urbano,
  JHEP {\bf 1709}, 010 (2017)
  doi:10.1007/JHEP09(2017)010
  [arXiv:1704.05438 [hep-ph]].

\bibitem{Amhis:2016xyh} 
  Y.~Amhis {\it et al.} [HFLAV Collaboration],
  Eur.\ Phys.\ J.\ C {\bf 77}, no. 12, 895 (2017)
  doi:10.1140/epjc/s10052-017-5058-4
  [arXiv:1612.07233 [hep-ex]].


\bibitem{DiLuzio:2017chi} 
  L.~Di Luzio and M.~Nardecchia,
  Eur.\ Phys.\ J.\ C {\bf 77}, no. 8, 536 (2017)
  doi:10.1140/epjc/s10052-017-5118-9
  [arXiv:1706.01868 [hep-ph]].


\bibitem{Freytsis:2015qca} 
  M.~Freytsis, Z.~Ligeti and J.~T.~Ruderman,
  Phys.\ Rev.\ D {\bf 92}, no. 5, 054018 (2015)
  doi:10.1103/PhysRevD.92.054018
  [arXiv:1506.08896 [hep-ph]].


\bibitem{Alonso:2015sja} 
  R.~Alonso, B.~Grinstein and J.~Martin Camalich,
  JHEP {\bf 1510}, 184 (2015)
  doi:10.1007/JHEP10(2015)184
  [arXiv:1505.05164 [hep-ph]].


\bibitem{Greljo:2015mma} 
  A.~Greljo, G.~Isidori and D.~Marzocca,
  JHEP {\bf 1507}, 142 (2015)
  doi:10.1007/JHEP07(2015)142
  [arXiv:1506.01705 [hep-ph]].


\bibitem{Calibbi:2015kma} 
  L.~Calibbi, A.~Crivellin and T.~Ota,
  Phys.\ Rev.\ Lett.\  {\bf 115}, 181801 (2015)
  doi:10.1103/PhysRevLett.115.181801
  [arXiv:1506.02661 [hep-ph]].


\bibitem{Crivellin:2017zlb} 
  A.~Crivellin, D.~Muller and T.~Ota,
  JHEP {\bf 1709}, 040 (2017)
  doi:10.1007/JHEP09(2017)040
  [arXiv:1703.09226 [hep-ph]].

\bibitem{Alonso:2016oyd} 
  R.~Alonso, B.~Grinstein and J.~Martin Camalich,
  Phys.\ Rev.\ Lett.\  {\bf 118}, no. 8, 081802 (2017)
  doi:10.1103/PhysRevLett.118.081802
  [arXiv:1611.06676 [hep-ph]].


\bibitem{Feruglio:2016gvd} 
  F.~Feruglio, P.~Paradisi and A.~Pattori,
  Phys.\ Rev.\ Lett.\  {\bf 118}, no. 1, 011801 (2017)
  doi:10.1103/PhysRevLett.118.011801
  [arXiv:1606.00524 [hep-ph]].


\bibitem{Feruglio:2017rjo} 
  F.~Feruglio, P.~Paradisi and A.~Pattori,
  JHEP {\bf 1709}, 061 (2017)
  doi:10.1007/JHEP09(2017)061
  [arXiv:1705.00929 [hep-ph]].


\bibitem{Faroughy:2016osc} 
  D.~A.~Faroughy, A.~Greljo and J.~F.~Kamenik,
  Phys.\ Lett.\ B {\bf 764}, 126 (2017)
  doi:10.1016/j.physletb.2016.11.011
  [arXiv:1609.07138 [hep-ph]].


\bibitem{Aad:2015caa} 
  G.~Aad {\it et al.} [ATLAS Collaboration],
  Eur.\ Phys.\ J.\ C {\bf 76}, no. 1, 5 (2016)
  doi:10.1140/epjc/s10052-015-3823-9
  [arXiv:1508.04735 [hep-ex]].


\bibitem{Sirunyan:2017yrk} 
  A.~M.~Sirunyan {\it et al.} [CMS Collaboration],
  JHEP {\bf 1707}, 121 (2017)
  doi:10.1007/JHEP07(2017)121
  [arXiv:1703.03995 [hep-ex]].


\bibitem{Dorsner:2017ufx} 
  I.~Dor\v sner, S.~Fajfer, D.~A.~Faroughy and N.~Ko\v snik,
  JHEP {\bf 1710}, 188 (2017)
  doi:10.1007/JHEP10(2017)188
  [arXiv:1706.07779 [hep-ph]].


\bibitem{Greljo:2017vvb} 
  A.~Greljo and D.~Marzocca,
  Eur.\ Phys.\ J.\ C {\bf 77}, no. 8, 548 (2017)
  doi:10.1140/epjc/s10052-017-5119-8
  [arXiv:1704.09015 [hep-ph]].


\bibitem{Allanach:2017bta} 
  B.~C.~Allanach, B.~Gripaios and T.~You,
  arXiv:1710.06363 [hep-ph].

\end{thebibliography}
\end{document}